\RequirePackage{fix-cm}
\documentclass[smallextended]{svjour3}       % onecolumn (second format)
\smartqed  % flush right qed marks, e.g. at end of proof
\usepackage{graphicx}
\def\beq{\begin{equation}}
\def\eeq{\end{equation}}
\def\bea{\begin{eqnarray}}
\def\eea{\end{eqnarray}}
\def\a{\alpha}
\def\b{\beta}
\def\r{\rho}

\def\m{\mu}
\def\n{\nu}
\def\h{\eta}
\def\s{\sigma}

\def\l{\lambda}

\def\G{\Gamma}
\def\F{\Phi}
\def\p{\pi}
\def\hh{\tilde{h}}

\def\da{\partial^2}
\def\aa{\partial}

\def\2{\;\;}
\def\4{\;\;\;\;}

\begin{document}

\title{Novel Ansatzes and Scalar Quantities in Gravito-Electromagnetism}


\author{A. Bakopoulos  \and P. Kanti }


\institute{A. Bakopoulos \at
              Division of Theoretical Physics, Department of Physics,
University of Ioannina, Ioannina GR-45110, Greece \\
              \email{abakop@cc.uoi.gr}           %  \\
           \and
           P. Kanti \at
              Division of Theoretical Physics, Department of Physics,
University of Ioannina, Ioannina GR-45110, Greece\\
              Tel.: +30-26510-08486\\
              Fax: +30-26510-08698\\
              \email{pkanti@cc.uoi.gr}
}

\date{Received: date / Accepted: date}

\maketitle

\begin{abstract}
In this work, we focus on the theory of Gravito-Electromagnetism (GEM) -- the theory
that describes the dynamics of the gravitational field in terms of quantities met
in Electromagnetism -- and we propose two novel forms of metric perturbations.
The first one is a generalisation of the traditional GEM ansatz, and succeeds in
reproducing the whole set of Maxwell's equations even for a dynamical vector
potential $\vec{A}$. The second form, the so-called alternative ansatz, goes
beyond that leading to an expression for the Lorentz force that matches the
one of Electromagnetism and is free of additional terms even for a dynamical
scalar potential $\Phi$. 
%In both cases, the theory is also free from any
%unphysical constraints that, in the case of the traditional analysis, follow
%from the spatial components of the field equations and gauge conditions.
In the context of the linearised theory, we then search for scalar invariant
quantities in analogy to Electromagnetism. We define three novel, 3rd-rank
gravitational tensors, and demonstrate that the last two can be employed to
construct scalar quantities that succeed in giving results very similar
to those found in Electromagnetism. Finally, the gauge invariance of the
linearised gravitational theory is studied, and shown to lead to the gauge
invariance of the GEM fields $\vec{E}$ and $\vec{B}$ for a general configuration
of the arbitrary vector involved in the coordinate transformations.

\end{abstract}

%%%%%%%%%%%%%%%%%%%%%%%%%%%%%%%%%%%%%%%%%%%%%%%%%%%%%%%%%%%%%%%%%%%%%%

\section{Introduction}
\label{Intro}

Of all the forces in nature, the gravitational force has proven to be the
most resistant to being incorporated into a common framework that would
unify all possible interactions. Whereas the electromagnetic, weak and
strong interactions have all been successfully described by gauge field
theories, gravity is accurately described by Einstein's General Theory of
Relativity, a mathematical theory based on tensors rather than on four-vectors.
An intrinsic difference arises regarding the spin of the fundamental degrees of
freedom in each case: while all gauge bosons have spin one, gravitons have
spin two; this inevitably affects the type of symmetry present in each theory,
and eventually determines the mathematical framework that describes it best.

At the same time, gravity could not be more similar to Electromagnetism (EM)
at the classical level. The similarity of the equations obeyed by the
Newtonian and Coulomb potential was noticed centuries ago, and this analogy
was re-inforced after the discovery of the Lense-Thirring effect \cite{LT}
where the angular momentum of a rotating body may be interpreted at large distances
as a gravitational `magnetic' field. All these cultivated the impression that
the unification of gravity with electromagnetism would be straightforward
and imminent. Numerous attempts have therefore appeared in the literature
over a century-long period including the Kaluza-Klein theory \cite{KK}, the string
and M-theory \cite{Witten}\cite{Duff}, the loop quantum gravity \cite{Ashtekar},
as well as a number of other geometric theories, classical or quantum, that
have attempted to connect gravity and electromagnetism 
\cite{Others-early}\cite{Others-recent} (for a more extensive list of
references including even earlier works, see \cite{Ramos-PhD}\cite{Bakopoulos}).
So far, the construction of a robust mathematical or geometrical theory, in the
context of which the unification of all forces could be realised, is still 
missing. 

A different perspective, that has attracted considerable attention over the years,
is the one adopted in the theory of Gravito-electromagnetism (GEM)
\cite{GEM-early}\cite{Damour}\cite{GEM-middle}\cite{static}\cite{non-static}\cite{Mashhoon}\cite{Costa}\cite{Ramos}\cite{Bouda}\cite{BK1}.
In this, the dynamics of the gravitational field is described in terms of quantities
met in electromagnetism. Within the GEM theory itself, there are mainly two different
approaches: the first one makes use of the decomposition of the Weyl and
Maxwell tensors in electric and magnetic parts \cite{GEM-middle}, while the
second focuses on the perturbed Einstein's equations and expresses the
gravitational perturbations in terms of a scalar and a vector GEM potential \cite{Mashhoon}.  
In the second approach, the GEM analogy emerges quite easily since the
perturbed field equations reduce to a set of Maxwell-like equations for the GEM
potentials in a straightforward way; the first approach, on the other hand,
relies on finding a set of equations for the electromagnetic parts of the Weyl
and Maxwell tensors that resemble the ones in Electromagnetism, a task that
is significantly more challenging.

The second approach is naturally of limited validity due to the weak-field
approximation formalism it uses; moreover, a particular expression for the
form of gravitational perturbations needs to be employed. The first approach
is valid for arbitrary gravitational fields, and it is thus exact and covariant.
Although one expects that a physical connection exists between the two approaches,
such a connection is not yet clear.
The exact approach, although more robust in a mathematical way, is also more
complex with not so obvious physical interpretation. The linear approach,
although of a limited validity, is simpler and provides a natural framework for
gravitational phenomena that happen far away from us. In the absence of the
connection between them, we consider these two approaches as complementary
that provide important information for the nature of gravitational field and its
analogy with Electromagnetism, information that may in the future be used for
the unification of all forces. 

In this work, we focus on the second, linear approach of the GEM theory, and attempt
to provide a remedy for a number of weak points present in the traditional
GEM analysis. To start with, a debate exists in the literature on whether the
GEM analogy holds only for static \cite{static} or for dynamical \cite{non-static}
backgrounds. In a previous work of ours \cite{BK1}, we performed a comprehensive
analysis of the set of equations that follow from the linearised field equations for
the so-called traditional ansatz for the metric perturbations \cite{Mashhoon}.
Our results demonstrated that these equations reduce to Maxwell's equations
only under the assumption of a static vector potential $\vec{A}$ - in fact, the
staticity of $\vec{A}$ was dictated by the neglected spatial component of the
transverse gauge condition, and was thus an intrinsic feature of the theory.
In addition, the geodesics equation, that in the context of GEM reduces to a
form similar to that of the Lorentz force, was showed to contain additional terms
that may not be easily ignored unless the scalar potential $\Phi$ is also static. 
We also showed that the spatial components of the field equations, that are also
often ignored in the literature, carry important pieces of information:
at times, they may impose unphysical or over-restrictive constraints to the
fields or matter distribution of the theory, and therefore should be properly
taken into account. 

An important conclusion that followed from the analysis of \cite{BK1} was that
the form of the gravitational perturbations significantly affects the form of the
field equations, the expression for the Lorentz force and the form of any
additional constraints. In the present work, motivated by our previous findings,
we introduce two novel forms of metric perturbations. The first one, the
{\it generalised traditional ansatz}, is a generalisation of the ansatz usually
employed in GEM that allows small but non-vanishing spatial components of
the metric perturbations $\tilde h_{\m\n}$. Although the $\tilde h_{ij}$
components are not involved in the derivation of Maxwell's equations, they affect
other important equations of the theory.  Indeed, we show that, in their presence,
the set of Maxwell's equations are exactly reproduced, and all terms involving
time-derivatives of the vector potential $\vec{A}$ are now restored. 
The second novel form of perturbations, the {\it alternative ansatz}, has as
its core idea the introduction of the scalar potential
$\Phi$ into the spatial components $\tilde h_{ij}$, too, of the metric perturbations.
Then, a cancellation of potentially harmful terms in the expression of the Lorentz
force leaves behind a minimal form identical, at first
approximation, to that in Electromagnetism.
%and with any additional corrections being negligible even in the relativistic limit.
%The additional components of the field equations are also rendered harmless,
In addition, no unphysical constraints on matter or field configurations arise in the
theory, and the set of Maxwell's equations is again restored for a dynamical vector
potential $\vec{A}$.
%with the only demand being the re-definition of the charge/matter density
%in the theory by a numerical factor. 

What is also important in establishing the extent of analogy between gravity and
EM is the construction of scalar invariant quantities in the context of GEM similar
to those in Electromagnetism. Previous attempts have appeared in the
literature before \cite{Costa} where either the Weyl or the Riemann tensor
is employed for this purpose. In fact, the ability to construct invariant quantities,
without relying on the linear approximation or on specific forms of the 
gravitational background, is one of the features that makes the exact approach
in GEM exact and covariant. This feature is missing from the linear approach,
therefore in this work we undertake the task of searching for scalar quantities
defined in the context of the linearised gravitational theory of GEM.
To this end, we define three novel 3rd-rank gravitational tensors in terms of
which we construct scalar quantities, and evaluate their role as analogues of
the scalar quantities of EM. These novel tensors are defined in a covariant way,
and their expressions may be computed for arbitrary backgrounds.
Here, we use the two novel ansatzes for the form of gravitational perturbations
introduced above, and demonstrate that, for two of those tensors, the results
resemble the EM scalar quantities. 

Finally, we turn our attention to the gauge invariance of the linearised gravitational
theory. It is well known that this may be interpreted as a gauge invariance
of the GEM fields $\vec{E}$ and $\vec{B}$ \cite{Mashhoon}.  However, this has
been demonstrated for a particular type of coordinate transformations. In
the context of the present analysis, we perform a comprehensive analysis,
and derive the most general constraints that the coordinate transformation
should obey in order for the gauge invariance of the GEM fields to hold.

The outline of our paper is as follows: in Section 2, we present the theoretical
framework of our analysis, review the most basic assumptions and equations
of GEM, and discuss the weak points of the traditional analysis. In Section 3,
we present the two novel forms of gravitational perturbations, and in each case
we derive the complete set of field equations, gauge condition and geodesics
equation. Then, in Section 4, we focus on the construction of scalar quantities
in terms of three novel gravitational tensors, and we compute their expressions
for both metric ansatzes. In Section 5, we address the topic of the gauge invariance,
derive the full set of constraints on the coordinate transformations and look for
the most general configuration. Finally, we present our conclusions in Section 6. 

%%%%%%%%%%%%%%%%%%%%%%%%%%%%%%%%%%%%%%%%%%%%%%%%%%%%%%%%%%%%%%%%%%%%
%
%%%%%%%%%%%%%%%%%%%%%%%%%%%%%%%%%%%%%%%%%%%%%%%%%%%%%%%%%%%%%%%%%%%%%

\section{The Theoretical Framework}
\setcounter{equation}{0}

Our theoretical framework will be the one developed in the context of the
linearised theory of Gravito-electromagnetism (GEM), therefore we start our
analysis by presenting briefly its basic assumptions and equations
\cite{Mashhoon}. The general metric tensor is assumed to be written as 
%%%%%%%%%%%%
\begin{equation}
g_{\mu\nu} (x^\mu) =\eta_{\mu\nu} + h_{\mu\nu}(x^\mu)\,,
\label{metric}
\end{equation}
%%%%%%%%%%%%%%
where $\eta_{\mu\nu}$ is the Minkowski metric of the flat
spacetime\footnote{Throughout this work, we will use the $(+1,-1,-1,-1)$ signature 
for the Minkowski tensor $\eta_{\mu\nu}$.} and $h_{\mu\nu}$ are the metric
perturbations. The latter are functions of $x^\mu=(ct, \vec{x})$ and are associated to
the presence of gravitating bodies. They are also assumed to obey the inequality
$|h_{\mu\nu}|\ll 1$, and therefore a linear-approximation analysis may be followed
for their study. 

By following a standard procedure \cite{Landau} and using a new form of metric perturbations,
defined through the relation
%%%%%%%%%%%%%%%
\begin{equation}\label{newh}
\hh_{\m\n}=h_{\m\n} - \frac{1}{2}\,\h_{\m\n}\,h\,,
\end{equation}
%%%%%%%%%%%%%%%
the perturbed Einstein's equations take the form
%%%%%%%%%%%%%%%
\begin{equation}\label{Einstein_new}
{\hh^{\a}}_{\2\m,\n\a}+{\hh^{\a}}_{\2\n,\m\a}-
\da \hh_{\m\n}-\h_{\m\n}\,{\hh^{\a\b}}_{\4,\a\b}=2k T_{\mu\nu}\,.
\end{equation}
%%%%%%%%%%%%%%%
In the above $k=8\pi G/c^4$, $\partial^2 \equiv \eta^{\mu\nu}\partial_\mu \partial\nu$
and $T_{\mu\nu}$ is the energy-momentum tensor. The latter is described by the
expression $T_{\mu\nu}=\rho\,u_\mu u_\nu$, where $\rho$ is the mass/charge density 
in the context of GEM and $u^\mu=(u^0,u^i)=(c,\vec{u})$ is the velocity of the source.

In the context of the traditional ansatz adopted in GEM, the components of the metric
perturbations $\tilde h_{\mu\nu}$ have the form \cite{Mashhoon}
%%%%%%%%%%%%
\beq 
\tilde h_{00}=\frac{4\Phi}{c^2}\,, \qquad \tilde h_{0i}=\frac{2A_i}{c^2}\,,
\qquad \tilde h_{ij}=\,\mathcal{O}(c^{-4})\,, \label{case1}
\eeq
%%%%%%%%%%%
where the scalar $\Phi(x^\mu)$ and vector $\vec{A}(x^\mu)$ functions are the
so-called gravito-electromagnetic potentials defined in analogy with electromagnetism.
In reality, $\Phi$ is the Newtonian gravitational potential while $\vec{A}$ is associated
to the angular-momentum vector, if existent, of the massive body. The spatial
components $\tilde h_{ij}$ of the perturbations are assumed to be negligible and are
thus ignored in the analysis. Working in the transverse gauge, i.e
$\hh^{\m\n}_{\4,\n}=0$, the $(00)$ and $(0i)$ components of the field equations
(\ref{Einstein_new}) take the form 
%%%%%%%%%%%%%%%%
\begin{equation}\label{poison_1}
\da \F\,=\,-4 \p  G \, \r,  \qquad 
\da \left(\frac{A^i}{2} \right)\,=\,-\frac{4\p G}{c}\,j^i,
\end{equation}
respectively, where $j^i\equiv \rho u^i$. By employing the following definitions of the
gravito-electromagnetic fields \cite{Mashhoon} in terms of the GEM potentials
%%%%%%%%%%%%%%%
\beq
\vec{E} \equiv -\frac{1}{c}\,\partial_t \left(\frac{\vec{A}}{2}\right) -\vec{\nabla} \Phi\,, 
\qquad \vec{B} \equiv \vec{\nabla} \times \left(\frac{\vec{A}}{2}\right)\,, \label{EB_GEM}
\eeq
%%%%%%%%%%%%%%%
Eqs. (\ref{poison_1}) adopt the form of two Maxwell-like equations for
$\vec{E}$ and $\vec{B}$, namely
%%% 
%%%%%%%%%%%%%%
\beq
\vec{\nabla} \cdot \vec{E} = 4\pi G \rho\,, \qquad 
\vec{\nabla} \times \vec{B}=\frac{1}{c}\,\partial_t \vec{E} + 
\frac{4\pi G}{c}\,\vec{j}\,. \label{Maxwell12_GEM}
\eeq
%%%%%%%%
The definitions of the fields Eqs. (\ref{EB_GEM}) may in turn take the form of the remaining
two Maxwell equations. In the above analysis, the supplementary assumption was also made
that the vector potential be static, i.e. $\partial_t A^i=0$.

Finally, the spatial components of the geodesics equation
%%%%%%%%%%%%%%%%%%%
\beq
\frac{d^2 x^\rho}{ds^2} + \Gamma^\rho_{\mu\nu}\,\frac{dx^\mu}{ds}
\frac{dx^\nu}{ds}=0\,, \label{geodesics-GEM}
\eeq
%%%%%%%%%%%%%%%%%%%
in the same linear approximation, and in the non-relativistic limit where 
$ds^2 \simeq c^2 dt^2$, may be collectively written as \cite{Mashhoon}
 %%%%%
\begin{equation}\label{lorentz-force1}
\ddot{x}^{\,i}=\,E^i +\frac{2}{c}\,F^{ij}u_j\,,
\end{equation}
%%%%%%%
where $F_{ij} \equiv \partial_i A_j -\partial_j A_i$. The above has the form of the Lorentz
equation of electromagnetism, however, only under the additional assumption that
the scalar potential is also time-independent, $\partial_t \Phi=0$.

In the previous work of ours \cite{BK1}, the above analysis was repeated without
the imposition of the gauge condition in order to investigate the reason for the 
required staticity of the GEM potentials and the role, if any, of the gauge condition
in this. Assuming again the ansatz (\ref{case1}) for the metric perturbations, we
found that the field equations (\ref{Einstein_new}) take in fact the following forms:
%%%%%%%%%%%%%%%%
\beq
\nabla^2\,\Phi=4 \pi G \rho\,,
\label{PoissonGEM}
\eeq
%%%%%%%%%%%%%%%
for $(\mu,\nu)=(0,0)$, and 
%%%%%%%%%%%%%%%%
\beq
\vec{\nabla}\,\left[\vec{\nabla} \cdot \left(\frac{\vec{A}}{2}\right) + 
\frac{1}{c}\,\partial_t \Phi\right] 
-\nabla^2 \left(\frac{\vec{A}}{2}\right) = \frac{4\pi G}{c}\,\rho \,\vec{u}\,,
\label{4MaxwellGEM}
\eeq
%%%%%%%%%%%%%%%%
for $(\mu,\nu)=(0,i)$. Equation (\ref{PoissonGEM}), the analogue of Poisson's law,
together with Eq. (\ref{4MaxwellGEM}) take indeed the form of the Maxwell's equations
(\ref{Maxwell12_GEM}) but only under the assumption of time-independence of the vector
GEM potential, since the anticipated terms involving $\partial_t A^i$ are missing. It is
therefore the need for the recovery of the analogy between GEM and electromagnetism
that demands the staticity of $\vec{A}$ in \cite{Mashhoon} and not the imposition
{\it per s\'e} of the gauge condition. Nevertheless, there is an underlying connection
since the ($\mu=i$)-components of the gauge condition $\hh^{\m\n}_{\4,\n}=0$ in
fact reduce to the constraint 
%%%%%%%%%%%%%%%%%
\begin{equation}\label{add_gauge1}
\frac{1}{c}\,\partial_t \vec{A}=0\,,
\end{equation}
%%%%%%%%%%%%%
that consistently complements the aforementioned requirement of a static vector
potential (the role of this constraint in the context of GEM was also studied in
\cite{GEM-early}f, \cite{GEM-middle}a,g  and \cite{Costa}).
 
Due to the tensorial structure of gravity, the field
equations (\ref{Einstein_new}) also yield additional constraints arising from the 
$(\mu,\nu)=(i,j)$ components -- except in a few analyses \cite{Costa}, these were
largely ignored in the literature. The diagonal and off-diagonal spatial components
were found in \cite{BK1} to have the form 
%%%%%%%%%%%%%%
\beq
\partial_t^2 \Phi =- \frac{\pi}{3} \rho\,|\vec{u}|^2\,, \qquad
\partial_0 \left(\partial_i A^j +\partial_j A^i\right)= 8\pi G\rho 
\,\frac{u_i u_j}{c^2}\,, \label{off-diagonal-GEM}
\eeq
%%%%%%%%%%%%%%%
respectively. The first of the above equations demands an isotropic distribution of sources 
and restricts the magnitude of $\partial_t^2 \Phi$ whereas the second dictates that, for
a time-independent vector potential, as demanded above, the source should be static
($u^i=0$) or its motion one-directional ($u^i u^j=0$). 

By repeating finally the derivation of the Lorentz equation from the geo\-desics
equation (\ref{geodesics-GEM}), we found that its complete form
is given by the expression \cite{BK1}
%%%%%%%%%%%%%%%
\begin{equation}
m\,\vec{a}=\vec{F}=m\,\vec{E}\left(1+\frac{|\vec{u}|^2}{c^2}\right)
+\frac{4m}{c}\,\vec{u}\times \vec{B} +
2m\left[\frac{\vec{u}}{c}\,\frac{\partial_t \Phi}{c} -
\frac{\vec{u}}{c}\,\left(\frac{\vec{u}}{c} \cdot \vec{E}\right)\right],
\label{Lorentz-GEM}
\end{equation}
%%%%%%%%%%%%%%
where, in accordance to the aforementioned discussion, we have already set
$\partial_t {\vec A}=0$. Still, the above expression is an extended one compared
to Eq. (\ref{lorentz-force1}) that appeared in \cite{Mashhoon}. The two additional
terms proportional to $u^2/c^2$ may be indeed safely ignored in the non-relativistic
limit.  However, the additional term proportional to the combination
$\vec{u}\,\partial_t \Phi/c^2$ is not equally suppressed and thus should be taken
into account -- unless the scalar potential is also time-independent, as assumed
in \cite{Mashhoon}. But although, in this case, this last term that spoils the
analogy with the electromagnetism in the expression of the Lorentz force indeed drops out,
a new problem arises: through the first of Eqs. (\ref{off-diagonal-GEM}), we are forced to restrict
our analysis only either to static distributions
of matter ($u=0$), or in pure vacuum ($\rho=0$); both requirements considerably
restrict the physical importance of the achieved analogy with electromagnetism
(for a more extended analysis on the link between the time-independence of the
GEM potentials and the analogy between gravity and Electromagnetism, see
\cite{Costa}).

%%%%%%%%%%%%%%%%%%%%%%%%%%%%%%%%%%%%%%%%%%%%%%%%%%%%%%

\section{Novel Ansatzes for the metric perturbations}
\setcounter{equation}{0}

In the previous section, we have demonstrated that, in the context of the traditional
GEM ansatz for the metric perturbations, a static vector potential $\vec{A}$
is necessary for the field equations to reduce to a set of Maxwell-like equations,
and that only for a static scalar potential $\Phi$ the form of the Lorentz force is
restored in the
non-relativistic limit. In this section, we will consider two alternative, more generalised
ansatzes for the metric perturbations, and investigate whether these two problems can be
simultaneously resolved without the imposition of the time-independence of the GEM fields. 

In both ansatzes, to be studied in the next two subsections, the metric perturbations
$\tilde h_{\mu\nu}$ will be assumed to have a form similar to
the one given in Eq. (\ref{case1}) but with the spatial components $\tilde h_{ij}$ not
being necessarily negligible. To justify this, let us derive the constraints that follow
from the transverse gauge condition $\tilde h^{\mu\nu}_{\4, \nu}=0$ while taking into
account the presence of the $\tilde h_{ij}$ components: while for $\mu=0$, we recover 
the usual Lorentz condition 
%%%%%%%%%%%%
\begin{equation}\label{gauge0-GEM}
\frac{1}{c}\,\aa_t \F + \aa_i \left(\frac{A^i}{2}  \right)=0\,,
\end{equation}
%%%%%%%%%%%%%%
for $\mu=i$, we now obtain the constraint
%%%%%%%%%%%%%
\beq
\frac{2}{c^3}\,\partial_t A^i+ \partial_j \tilde h^{ij}=0\,.
\label{lorentz_extra}
\eeq
%%%%%%%%%%%%%%%
We therefore conclude that the time-dependence of the vector GEM potential $\vec{A}$
is directly related to the spatial components of the metric perturbations $\tilde h_{ij}$.
If these components are altogether ignored in the analysis, as in the 
traditional GEM ansatz \cite{Mashhoon}, the time-dependence of $\vec{A}$
is automatically eliminated (on this, see also \cite{Bouda}).

%%%%%%%%%%%%%%%%%%%%%%%%%%%%%%%%%%%%%%%%

\subsection{A generalised form of the traditional ansatz}

Here, we extend the traditional form assumed for the metric perturbations in the context of
GEM \cite{Mashhoon}, and write the following generalised form
%%%%%%
\beq 
\tilde h_{00}=\frac{4\Phi}{c^2}\,, \qquad \tilde h_{0i}=\frac{2A_i}{c^2}\,,
\qquad \tilde h_{ij}=\frac{2\l}{c^4}\,\h_{ij}\,+\,\frac{2}{c^4}\,d_{ij}\,.
\label{case_general}
\eeq
%%%%%%
The spatial components $\tilde h_{ij}$ are assumed to be small but nevertheless
non-vanishing; in accordance to the aforementioned
discussion, we expect that this will restore
the time-dependence of the vector potential $\vec{A}$ in the analysis. In addition, their
expression may be decomposed, in the most general case, into two parts: one involving
a scalar function $\l$ and one that is proportional to a second-rank symmetric tensor
$d_{ij}$ -- the latter will be taken to be traceless since this choice simplifies significantly
the subsequent analysis and results. 

Let us first address the question of the form of the field equations. Considering
the $(00)$ component of the perturbed Einstein's equations (\ref{Einstein_new})
and employing the gauge condition (\ref{lorentz_extra}), we derive the equation
%%%%%%%%%%%%%%%%%
\beq
-\delta^{ij} \partial_i \partial_j \Phi -
\frac{1}{2c}\,\partial_i (\partial_t A^i) =- 4\pi G \rho\,.
\label{comp00-v1}
\eeq
%%%%%%%%%%%%%%%%%%%%
Defining the GEM field $\vec{E}$ as in Eq. (\ref{EB_GEM}), the above adopts a form
identical to the first Maxwell-like equation (\ref{Maxwell12_GEM}). Note that, as expected,
the presence of the $\tilde h_{ij}$ component has restored the term proportional to
$(\partial_t A^i)$ in the above equation through the improved gauge condition
(\ref{lorentz_extra}). Moving to the $(0i)$ component of Einstein's equations, this is
now found to be
%%%%%%%%%%%%
\beq
\frac{1}{2}\left[\partial_i (\partial_j A^j) - \delta^{kj} \partial_k \partial_j A^i\right] -
\frac{1}{c}\,\partial_t \left(-\partial_i \Phi -\frac{1}{2c}\,\partial_t A^i\right)
=-\frac{4\pi G}{c}\,\rho u^i\,,
\eeq
%%%%%%%%%%%%
after using again the gauge condition (\ref{lorentz_extra}).
If we use the definition of the GEM field $\vec{B}$, as this is given in Eq. (\ref{EB_GEM}),
the above equation again reduces to the second Maxwell-like equation (\ref{Maxwell12_GEM})
with the $\partial_t A^i$ term present as expected. The remaining two Maxwell-like equations
follow without a problem.
%%%%%%%%%%%%%%%

Finally, the spatial components $(ij)$ of the Einstein's equations (\ref{Einstein_new}), after
using both gauge conditions (\ref{gauge0-GEM}) and (\ref{lorentz_extra}),
lead to the additional relation
%%%%%%%%%%%%%%%%%
\begin{equation}\label{adeq6}
\da \,\tilde h ^{ij}=\frac{2}{c^4}\,\partial^2 (\l\,\eta^{ij} +d^{ij})=
-\frac{16 \p G}{c^4}\,j^i\,u^j\,.
\end{equation}
%%%%%%%%%%%%%
We observe that, in this case, the additional components -- instead of imposing
over-restrictive constraints to the fields or charge distribution, as in the traditional
GEM ansatz  [see Eqs. (\ref{off-diagonal-GEM})] -- 
simply relates the scalar $\lambda$ and tensor $d_{ij}$ potentials to the distribution
of matter/charge of the system. 

We now proceed to the expression of the Lorentz force. The geodesics equation 
(\ref{geodesics-GEM}) involves the Christoffel symbols which, in the linear approximation,
assume the form
%%%%%%%%%%%%%%%
\begin{equation}\label{Christoffel}
\Gamma^{\alpha}_{\mu\nu}=\frac{1}{2}\,\eta^{\alpha\rho}\left(
h_{\mu\rho,\nu}+h_{\nu\rho,\mu}-h_{\mu\nu,\rho}\right).
\end{equation}
%%%%%%%%%%%%%%%%%
We therefore need the original perturbations $h_{\mu\nu}$: by contracting
Eq. (\ref{newh}) by $\eta^{\mu\nu}$, one finds that $h=-\tilde h$; then, the
inverse relation between the original and the new perturbations may be written as 
%%%%%%%%%%%%%%%%%
\beq
h_{\mu\nu}= \tilde h_{\mu\nu} -\frac{1}{2}\,\eta_{\mu\nu}\,\tilde{h}\,.
\label{inversehnew}
\eeq
%%%%%%%%%%%%%%%%%%%
For the perturbations considered here, given by Eq. (\ref{case_general}), we find that
%%%%%%%%%%%%%
\beq
\tilde{h}=\frac{4\Phi}{c^2}+\frac{6\lambda}{c^4}\,,
\label{trace_h}
\eeq
%%%%%%%%%%%%
where we have used the fact that $\eta^{ij}d_{ij}=0$. The above, together with
Eq. (\ref{inversehnew}), leads to the original perturbations $h_{\mu\nu}$, or equivalently
to the following spacetime line-element through Eq. (\ref{metric}):
%%%%%%%%%%%%%%%%%%
\bea
\hspace*{-2.5cm} && ds^2=c^2\left(1+\frac{2\Phi}{c^2}-\frac{3\lambda}{c^4}\right)dt^2-
\frac{4}{c}\,(\vec{A} \cdot d\vec{x})\,dt \nonumber \\[1mm]
&& \hspace*{2.5cm}- \left(1-\frac{2\Phi}{c^2}-\frac{\lambda}{c^4}\right)\delta_{ij} dx^i dx^j +
\frac{2d_{ij}}{c^4}\,dx^i dx^j\,.
\label{line-element-general}
\eea
%%%%%%%%%%%%%%%%%%

Employing the expression of the Christoffel symbols (\ref{Christoffel}) and of the initial
perturbations $h_{\mu\nu}$ from the line-element above, we easily
find the components of $\Gamma^{\alpha}_{\mu\nu}$ - these are listed in
Eq. (\ref{Christ_trad}) of the Appendix A. Substituting these in the spatial components
of the geodesics equation (\ref{geodesics-GEM}), we finally obtain 
%%%%%%%%%%%%%%%%%%%
\bea
\frac{d^2 x^i}{dt^2} &=& E^i+\frac{2}{c}\,F^{ij} u_j 
+\frac{1}{c^2}\left(2u^i \partial_t \Phi + 2u^i u^j \partial_j \Phi - u^2 \partial^i \Phi\right)
\nonumber \\[1mm]
&+& \frac{3}{2c^2}\,\partial_j d^{ij} +\frac{1}{2c^4}\left(2u^i \partial_t \lambda + 
2u^i u^j \partial_j \lambda - u^2 \partial^i \lambda\right)
\nonumber \\[1mm]
&-& \frac{1}{c^4}\,\left(2 u^j \partial_t d^i_{\,\, j}+
2u^k u^j \partial_j d^i_{\,\, k} - u^k u^j \partial^i d_{jk}\right)\,.
\label{Lorentz-general}
\eea
%%%%%%%%%%%%%%%%%%
Note that above we have used the definition $F_{ij} \equiv \partial_i A_j -\partial_j A_i$.
The above expression provides a generalised form for the Lorentz force in the context
of GEM. By setting $d_{ij}=0$ and $\lambda=0$, we recover Eq. (\ref{Lorentz-GEM}) 
for the traditional ansatz (\ref{case1}) with the $u^i \partial_t \Phi/c^2$ term having
a significant contribution, even in the non-relativistic limit. In the presence of the
$d_{ij}$ and $\lambda$ potentials, additional terms arise in the expression of the
Lorentz force: although most come with a coefficient of ${\cal O}(1/c^4)$ and are
thus significantly suppressed, the term $\partial_j d^{ij}/c^2$ cannot again be
easily ignored even in the non-relativistic limit. 

We thus conclude that the generalised form of the metric perturbations (\ref{case_general})
employed in this subsection, although it exactly recovers all Maxwell-like equations 
without demanding the staticity of the vector potential $\vec{A}$, it cannot avoid the
presence of corrections in the expression of the Lorentz force. Before, however,
hastening to reject this ansatz, we note the similarity with which the scalar potentials
$\Phi$ and $\lambda$ appear in the expression of the Christoffel symbols and of the
Lorentz force. In the next subsection, we will therefore investigate whether a more
elaborate form of the metric perturbations $\tilde h_{\mu\nu}$, in which the $\Phi$
and $\lambda$ potential are related, can ameliorate the problems in the expression
of the Lorentz force.

%%%%%%%%%%%%%%%%%%%%%%%%%%%%%%%%%%%%%%%%%%%%%%%%%%%%%%%%%%%%%%%%%%%%%

\subsection{An alternative form of the metric perturbations}

A form of the metric perturbations, in which the $(00)$ and $(ij)$ components of
$\tilde h_{\mu\nu}$ are related, was employed also in our earlier work \cite{BK1}.
In there, the following relation was assumed to hold between the aforementioned
components of the metric perturbations
%%%%%%%%%%%%%%%
\beq
\tilde h_{ij}=-\tilde h_{00}\,\eta_{ij}\,.
\label{new-relation}
\eeq
%%%%%%%%%%%%%%%
Although traditionally it is only the $\tilde h_{00}$ component that is associated
with the scalar potential $\Phi$, the above assumption introduced a dependence
on $\Phi$ also in the spatial components of the metric perturbations. 
The above relation was shown to lead to the cancellation of all the additional terms
appearing in the expression of the Lorentz force leaving behind only the well-known
form of electromagnetism with the exact same coefficients. The weak point of this
ansatz was the fact that it was valid only in vacuum and, although it could successfully
describe the dynamics of the fields as they propagate, it failed to provide a 
robust framework for the study of the fields close to sources.

Guided by the above results, in the present analysis we will consider an alternative,
more extended, form of the metric perturbations, namely
%%%%%%
\beq 
\tilde h_{00}=\frac{\Phi}{c^2}\,, \qquad \tilde h_{0i}=\frac{A_i}{c^2}\,,
\qquad \tilde h_{ij}=-\gamma \frac{\Phi}{c^2}\,\h_{ij}\,+\,\frac{1}{c^4}\,d_{ij}\,.
\label{case_alter}
\eeq
%%%%%%
The $(00)$ and $(0i)$ components of $\tilde h_{\mu\nu}$, modulo numerical coefficients,
are identical to the ones in the previously considered ansatz (\ref{case_general}). The
$(ij)$ component involves again the symmetric, traceless, second-rank tensor $d_{ij}$
suppressed by a factor of $1/c^4$, as well as the scalar potential $\Phi$ suppressed
only by a factor of $1/c^2$ - the latter addition is necessary if the desired
cancellation of terms arising from the $\tilde h_{00}$ and $\tilde h_{ij}$ components
is to be realised. The constant coefficient $\gamma$, multiplying $\Phi$ in the
$\tilde h_{ij}$ component, will be determined by demanding the absence of corrections
in the expression of the Lorentz force. 

To this end, we need again the original perturbations $h_{\mu\nu}$ that appear in the
Christoffel symbols (\ref{Christoffel}). Employing Eq. (\ref{case_alter}), we find that
%%%%%%%%%%%%%
\beq
\tilde{h}=(1-3\gamma)\,\frac{\Phi}{c^2}\,.
\label{trace_h_alter}
\eeq
%%%%%%%%%%%%
By using again Eq. (\ref{inversehnew}), the spacetime line-element involving the original
perturbations $h_{\mu\nu}$ now takes the form
%%%%%%%%%%%%%%%%%%
\bea
\hspace*{-2.5cm} && ds^2=c^2\left[1+(1+3\gamma)\,\frac{\Phi}{2c^2}\right]dt^2-
\frac{2}{c}\,(\vec{A} \cdot d\vec{x})\,dt \nonumber \\[1mm]
&& \hspace*{2.5cm}- \left[1-(1-\gamma)\,\frac{\Phi}{2c^2}\right]\delta_{ij} dx^i dx^j +
\frac{d_{ij}}{c^4}\,dx^i dx^j\,.
\label{line-element-alter}
\eea
%%%%%%%%%%%%%%%%%%
The above leads to the components of the Christoffel symbols (\ref{Christoffel})
which are now listed in Eq. (\ref{Christ_alter}) of the Appendix A. 
Substituting again these components into the spatial components of
Eq. (\ref{geodesics-GEM}), we find 
the following result for the Lorentz force
%%%%%%%%%%%%%%%%%%%
\bea
\frac{d^2 x^i}{dt^2} &=&-\frac{(1+3\gamma)}{4}\partial^i \Phi -\frac{1}{c}\partial_t A^i
+ \frac{1}{c}\,F^{ij} u_j \nonumber \\[1mm]
&+&\frac{1}{2c^2}\,(1-\gamma)\,\left(u^i \partial_t \Phi +
u^i u^j \partial_j \Phi - u^2 \partial^i \Phi\right) \nonumber \\[1mm]
&-& \frac{1}{2c^4}\,\left(2 u^j \partial_t d^i_{\,\, j}+
2u^k u^j \partial_j d^i_{\,\, k} - u^k u^j \partial^i d_{jk}\right)\,.
\eea
%%%%%%%%%%%%%%%%%%
For $\gamma=0$, the analysis reduces again to that of the traditional ansatz 
(\ref{case1}); however, for $\gamma=1$, all terms of the order
${\cal O}(1/c^2)$ are eliminated leaving behind the minimal expression
%%%%%%%%%%%%%%%
\begin{equation}
m\,\vec{a}=\vec{F}=m\,\vec{E}+\frac{m}{c}\,\vec{u}\times \vec{B}
- \frac{m u^j}{2c^4}\,\left(2 \partial_t d^i_{\,\, j}+
2u^k \partial_j d^i_{\,\, k} - u^k \partial^i d_{jk}\right),
\label{Lorentz-alter}
\end{equation}
%%%%%%%%%%%%%%
under the usual definitions for the GEM fields 
%%%%%%%%%%%%%%%
\beq
\vec{E} \equiv -\frac{1}{c}\,\partial_t \vec{A} -\vec{\nabla} \Phi\,, 
\qquad \vec{B} \equiv \vec{\nabla} \times \vec{A}\,. \label{EB_alter2}
\eeq
%%%%%%%%%%%%%%%
The only additional terms in the expression of the Lorentz force are of
${\cal O}(1/c^4)$ that can be safely ignored even for large velocities.
In this respect, the result matches the one produced in \cite{BK1} where
the $d_{ij}$ term in the metric perturbations was altogether ignored --
as we will see, the inclusion of this term in the context of the present analysis,
although of small magnitude, will ensure again the consistency of the set of field
equations and their validity even for non-vacuum configurations. 

Next, we turn to the constraints that follow from the transverse gauge condition,
$\tilde h^{\mu\nu}_{\4,\nu}=0$. The temporal component $\mu=0$ gives the
exact Lorentz condition of electromagnetism, namely:
%%%%%%%%
\begin{equation}\label{lor6}
\frac{1}{c}\,\aa_t \Phi +\aa_i\,A^i\,=\,0. 
\end{equation}
%%%%%%%%
On the other hand, the spatial component $\m=i$ gives the following equation
%%%%%%%%
\begin{equation}\label{extra_v0} 
-\frac{1}{c}\,\aa_t A^i -\aa_i\F\,=\,E^i\,=\,\frac{1}{c^2} \,\aa_jd^{ij}\,,
\end{equation}
%%%%%%%%%
that relates the gravitoelectric field with the divergence of the tensor potential $d_{ij}$.
Note that, in the absence of the $d_{ij}$ term, the above condition would demand
the vanishing of the GEM field $\vec{E}$ and, through the field equations, the
vanishing of the source -- that was the problem encountered in the analysis
of \cite{BK1} that restricted the validity of the results only in vacuum configurations
or far away from the source. 
%Also, we should stress that the $\Phi$ term on
%the left-hand-side of  Eq. (\ref{add6}) comes, not from the $\tilde h_{00}$ component,
%but from the $\tilde h_{ij}$ component in the ansatz (\ref{case_general}) with $\gamma=1$. 
%Had this term been absent, it would have been the time-derivative of the vector
%field associated with the spatial derivative of $d_{ij}$ as in Eq. (\ref{lorentz_extra});
%for vanishing $d_{ij}$, we would have to work again with static fields.
Here, as also in the previous subsection, the presence of the $d_{ij}$ potential changes
the extra component of the gauge condition from an unphysical constraint to a
supplementary equation that relates its spatial derivative to the GEM electric field.

Let us finally address the question of the form of the field equations. These follow quite
easily from the ones derived in the previous subsection by performing the changes
$\Phi \rightarrow \Phi/4$ and $A^i \rightarrow A^i/2$. Thus, the $(00)$ and 
$(0i)$ components of the perturbed Einstein's equations (\ref{Einstein_new}) are
found to be
%%%%%%%%%%%%%%%%%%
\beq
\da  \Phi  =\,- 16 \p G \,\r\,, \qquad \da A^i\,=\,-\frac{16 \p G}{c}\,j^i\,,
\label{comp00-0i-v2}
\eeq
%%%%%%%%%%%%%
respectively -- in the above, we have used both gauge conditions (\ref{lor6})-(\ref{extra_v0}).
We thus observe that these two components of the field equations adopt indeed
forms identical to the Maxwell equations (\ref{Maxwell12_GEM}) apart from a factor of
4 on their right-hand-sides. The above are supplemented by the spatial components
$(ij)$ of the Einstein's equations, given by
%%%%%%%%%%%%%%%%%
\begin{equation}\label{additional}
\da  \,d^{ij} =\,-8 \p G\left( j^i u^j + \r c^2 \h^{ij} \right),
\end{equation} 
%%%%%%%%%%%%%%%%%%
after having applied again the gauge conditions (\ref{lor6})-(\ref{extra_v0})
Also in this case, the additional components of the perturbed Einstein's
equations simply relate the tensor potential $d_{ij}$ to the distribution of
matter/charge of the system. 

The above results constitute a significant improvement compared to the ones that follow
in the case of the traditional GEM ansatz (\ref{case1}). Here, the time-dependence of
both GEM potentials is restored: a set of four Maxwell-like equations may be recovered
not only for static but also for
a dynamical vector potential $\vec{A}$, and the expression of the Lorentz force matches the
exact electromagnetic one without having to assume that the scalar potential $\Phi$ is
static, too. Also, as noted above, the field configurations are free from unphysical constraints
since the additional components serve to determine the tensor potential $d_{ij}$ whose
contribution to observable effects, such as the Lorentz force, remains always suppressed.

The relation (\ref{new-relation}), that is still respected by the ansatz (\ref{case_alter})
at order ${\cal O}(1/c^2)$, places the resulting gravitational background (\ref{line-element-alter})
in a class of spacetimes where the spatial metric is written in terms of the so-called
``conformally Cartesian'' coordinates. This class of spacetimes was introduced in the
seminal work of \cite{Damour} on relativistic celestial mechanics as the most appropriate
for the study of physical problems. The reason for this is the fact that for such spacetimes
the spatial part of the metric is flat. Indeed, one may readily observe that for $\gamma=1$,
for which the relation (\ref{new-relation}) is satisfied, the spatial part of the line-element
(\ref{line-element-alter}) is flat at order ${\cal O}(1/c^2)$ independently of the values of
the $\Phi$ and $\vec{A}$ potentials. In this way, the coordinate invariance that is inherent in
general relativity is reduced, without being over-restricted, to the appropriate level for
the study of physical problems in gravity.  

In the context of the present analysis, as well as in that of \cite{Damour}, the $\tilde h_{00}$
and $\tilde h_{ij}$ are allowed to differ by terms of order ${\cal O}(1/c^4)$ - this is the
term proportional to the $d_{ij}$ tensor in (\ref{case_alter}). In the absence of this term, the
spatial part of the metric would be flat at all orders: this was the case studied in 
\cite{BK1} where we encountered the problem of the vanishing of the source - indeed,
the exactly flat character of space did not allow for the presence of any sources.
In the presence of this term, though, we avoid any unphysical constraints, and the derived
equations describe the fields not only in vacuum but also in the presence of sources. 
As a result, our present analysis naturally improves also the one presented in \cite{BK1}.

The only shortcoming that destroys the
exact similarity to the corresponding formulae of electromagnetism is the numerical
coefficient of $16\pi$, instead of $4\pi$, on the right-hand-side of the Maxwell-like
equations. One could suggest as a possible remedy the redefinition of the GEM potentials
according to the rule
%%%%
\beq
\F \longrightarrow 4 \F^*, \qquad
\vec{A} \longrightarrow 4 \vec{A^*},
\eeq
%%%
where $\F^*$ and $\vec{A^*}$ are now the true potentials. Then, defining the
gravitoelectric and gravitomagnetic field as
%%%%%%%%%%%%%%%
\beq\label{ebdef}
\vec{E} \equiv - \frac{1}{c}\,\partial_t \vec{A^*} -
\vec{\nabla} \Phi ^* \,, 
\qquad \vec{B} \equiv \vec{\nabla} \times \vec{A^*}\,,
\eeq
%%%%%%%%%%%%%%%
one may easily see that Eqs. (\ref{comp00-0i-v2}) adopt the correct form.
However, in this case, the redefinition of the GEM fields would unavoidably introduce an
additional numerical factor of 4 in the expression of the Lorentz force (\ref{Lorentz-alter})
that would again destroy the exact similarity with electromagnetism. We may thus conclude
that although the perturbations ansatz employed here has come a long way in overcoming
major obstacles in the analogy between gravity and electromagnetism, an exact matching
between the corresponding field equations is still eluding us. An alternative perhaps suggestion
would be the redefinition of the matter/charge density according to the rule
$\rho_{e}=4\rho_{m}$, where $\rho_{m(e)}$ is the matter and corresponding charge density,
respectively: in this case, all the positive features of the perturbations ansatz employed
here would be retained, and the exact matching in the form of all equations between
gravity and electromagnetism would be achieved.

%%%%%%%%%%%%%%%%%%%%%%%%%%%%%%%%%%%%%%%%%%%%%%%%%%%%%%%%%%%%%%%%%%%%%

\section{Scalar Invariant Quantities in GEM}
\setcounter{equation}{0}

Pursuing further the analogy between GEM and electromagnetism, in this section we will
search for scalar quantities -- and thus invariant under coordinate transformations
-- constructed in the context of GEM and conveying information similar to that in
electromagnetism. As is well-known, in pure electromagnetism one may construct two
such invariant quantities: the inner product of the electromagnetic field strength tensor
%%%%%%%%%%%%%%%%
\beq
F^{\mu\nu}\,F_{\mu\nu}=2\,(B^2-E^2)\,,
\label{inner-Fmn}
\eeq
%%%%%%%%%%%%%%%%%
and its product with its dual tensor $\tilde F^{\mu\nu}$, namely
%%%%%%%%%%%%
\beq
\tilde F^{\mu\nu}\,F_{\mu\nu}= \frac{1}{2}\,\epsilon^{\mu\nu\rho\sigma} F_{\rho\sigma}
F_{\mu\nu}= 4\,\vec{E} \cdot \vec{B}\,.
\label{antisym-Fmn}
\eeq
%%%%%%%%%%%%%% 
Both quantities (\ref{inner-Fmn}) and (\ref{antisym-Fmn}) are also gauge invariant,
with the first one being of fundamental importance as it appears in the expression of both
the Lagrangian of the electromagnetic field and its energy-momentum tensor. 

Constructing similar quantities in the context of GEM is not an easy task. Although scalar
quantities may be easily constructed in terms of gravitational quantities, such as the
Riemann and the Ricci tensor, these involve second derivatives of the metric tensor and
therefore second derivatives of the GEM potentials $\Phi$ and $\vec A$. As a result, any
invariant quantity constructed in this way would unavoidably involve not the GEM fields
$\vec E$ and $\vec B$, as the ones in Eqs. (\ref{inner-Fmn}) and (\ref{antisym-Fmn}),
but their first derivatives. 

It is possible to construct invariants similar to those of electromagnetism by using
the Weyl or the Riemann tensor \cite{Costa}, however this approach deviates from the
one followed in the linear approach in GEM. Here, we will investigate whether it is
possible to construct scalar invariants by using also gravitational quantities but
in the context of the linearised approach followed so far in this work.

%%%%%%%%%%%%%%%%%%%%%%%%%%%%%%%%%%%%

\subsection{A generalised field-strength tensor}

In electromagnetism, the field-strength tensor $F^{\mu\nu}$ appears also in the 
relativistic field equations, namely
%%%%%%%%%%%%%%
\beq
\partial_\mu F^{\mu\nu}=4 \pi \,J^\nu\,.
\label{EM-field-eq}
\eeq
%%%%%%%%%%%%%%%
Motivated by this, we turn for guidance to the perturbed Einstein's equations 
(\ref{Einstein_new}). In the context of gravity, these can be alternatively written
as \cite{Bakopoulos}
%%%%%%%%%%%%%%%%%
\begin{equation} 
G_{\m\n}=\frac{1}{2}\,\aa^\a F_{\a\m\n}=k\,T_{\m\n}\,,
\label{Einstein-F-eq}
\end{equation}
%%%%%%%%%%%%%%%%%%%
where the tensor $F_{\a\m\n}$ is defined as
%%%%%
\begin{equation}\label{fdef}
F_{\a\m\n}= \aa_\m \hh_{\a\n} + \aa_\n \hh_{\a\m} - \aa_\a \hh_{\m\n} -
\h_{\m\n}\, \aa^\b \hh_{\a\b}\,.
\label{Famn}
\end{equation}
%%%%%%%%%
This new tensor is a third-rank tensor that is also symmetric in the last two indices,
therefore it looks distinctly different from the electromagnetic field strength tensor
$F_{\mu\nu}$. However, the similarity between Eqs. (\ref{EM-field-eq}) and
(\ref{Einstein-F-eq}) motivates us to consider the following scalar combination
%%%%%
\begin{equation}\label{fscaldef}
F=F^{\a\m\n}F_{\a\m\n}
\end{equation}
%%%%%%%
as a plausible analogue of the electromagnetic invariant (\ref{inner-Fmn}). In what follows,
we will compute the aforementioned scalar quantity for the two ansatzes (\ref{case_general})
and (\ref{case_alter}) for the metric perturbations. 

We start with the generalised form (\ref{case_general}) of the metric perturbations, that
succeeded in restoring the time dependence of the vector potential, and we compute the
components of the new tensor $F_{\a\m\n}$ (\ref{Famn}). Their explicit form is given in
Eq. (\ref{Famn_general}) in the Appendix B. Using those, the scalar quantity
$F$ can be explicitly written~\footnote{From this point onwards
and in order to simplify the analysis,
we will set $\lambda=0$; as is clear by looking at the components (\ref{Famn_general}),
all terms related to the $\lambda$ potential will be suppressed by, at least, a factor
of ${\cal O}(1/c^6)$ in the expression of the $F$ invariant quantity (\ref{fscaldef})
and thus will be negligible -- by using the components (\ref{Famn_general}),
the interested reader may compute the full expression of the $F$ invariant and check
that indeed the presence of the scalar potential $\lambda$ does not alter the results
in any significant way.} in the following way
%%%%
%\begin{equation}\label{fexp}
%F=F^{000}F_{000}+2F^{00i}F_{00i}+F^{0ij}F_{0ij}+F^{i00}F_{i00}+2F^{ij0}F_{ij0}+F^{ijk}F_{ijk}.
%\end{equation}
%%%%%%%%%%%%%%%
%%%%%
%%%%%
\beq \label{fscal-v1}
F = \frac{4}{c^4} \left[8(2 B^2 - E^2) -4 \left(\aa_i \F \right)^2
+\left( \aa_i A_j +\aa_j A_i \right)^2 + \left(\aa_i A^i\right)^2\right] + \tilde R\,,
\eeq
%%%%%%%%%%%%%%%
where $\tilde R$ stands for the sum of the sub-dominant terms
%%%%%%%%%%%
\bea
R&=& -\frac{16}{c^7}\,\partial_t d_{ij}\,(\partial^i A^j)
-\frac{4}{c^8}\,(\partial_j d_{ik} +\partial_k d_{ij}-\partial_i d_{jk})^2 \nonumber \\[1mm]
&+&
\frac{8}{c^8}\,(\partial_j d^{ij})\,(\partial^k d_{ik}) +\frac{12}{c^{10}}\,(\partial_t d_{ij})^2\,.
\label{R-general}
\eea
%%%%%%%%%%%%%
In the above expressions, both gauge conditions (\ref{gauge0-GEM}) and (\ref{lorentz_extra})
have been used: according to the discussion of Section 3.1, the implementation of the latter
is necessary for the successful restoration of Maxwell-like equations while the use of the former
significantly simplifies the result. The scalar quantity $F^{\a\m\n} F_{\a\m\n}$, as given in
Eq. (\ref{fscal-v1}), has the desired
quadratic dependence on the GEM fields $\vec{E}$ and $\vec{B}$, however, the exact
result is not satisfactory: the numerical coefficients are not the expected ones and,
more importantly, additional terms, of equal magnitude compared to $E^2$ and $B^2$,
arise in its expression. Finally, we note the presence of the symmetric combination 
$(\aa_i A_j +\aa_j A_i)$ which is the result of the symmetry in the last two indices of
$F_{\a\m\n}$, a symmetry that is of course absent in the electromagnetic tensor $F_{\m\n}$.

In order to investigate how much the result given in Eq. (\ref{fscal-v1}) depends on
the specific ansatz for the metric perturbations, we will now consider the ansatz (\ref{case_alter})
that has led to the most successfull expression of the Lorentz force. Making use of the
components of the generalised tensor $F_{\a\m\n}$, that are now given in 
Eq. (\ref{Famn_alter}) of the Appendix B, and applying both gauge conditions
(\ref{lor6}) and (\ref{extra_v0}), we obtain the following final result for the scalar
$F$
%%%%%
\begin{eqnarray}\label{fscal-v0-alter}
F &=& \frac{1}{c^4} \left[4\,(B^2 -E^2) -10 \left(\aa_i \F \right)^2
+\left( \aa_i A_j +\aa_j A_i \right)^2\right] + \tilde R\,,
\end{eqnarray}
%%%%%%%%%%%%%%%
where $\tilde R$ stands again for the sum of the sub-dominant terms that now has the form
%%%%%%%%%%%
\bea
\tilde R&=&\frac{6}{c^6}\,(\partial_t \Phi)^2 -\frac{4}{c^7}\,\partial_t d_{ij}\,(\partial^i A^j)
-\frac{1}{c^8}\,(\partial_j d_{ik} +\partial_k d_{ij}-\partial_i d_{jk})^2 \nonumber \\[1mm]
&+& \frac{2}{c^{10}}\,(\partial_t d_{ij})^2\,.
\label{R-alter}
\eea
%%%%%%%%%%%%%
The definitions for the GEM fields that were used in Eq. (\ref{fscal-v0-alter}) are the ones
given in Eq. (\ref{EB_alter2}). We note that the use of the alternative ansatz (\ref{case_alter})
has improved the expression for the scalar invariant $F$, yielding the desired combination
$(B^2-E^2)$, however the appearance of additional terms of equal magnitude, although
more restricted compared to the case of the ansatz (\ref{case_general}), still can not be avoided.

In the light of the above results, we conclude that the selection of the generalised field-strength
tensor $F^{\a\m\n}$ (\ref{fdef}) for the construction of invariant quantities in the
context of GEM, although well motivated due to the validity of Eq. (\ref{Einstein-F-eq}),
was not a completely successful one. As a result, we will not
attempt to construct here a second invariant of the form $\tilde{F}^{\a\m\n}F_{\a\m\n}$,
where $\tilde{F}^{\a\m\n}$ is a dual form of ${F}^{\a\m\n}$, as the analogue of the
electromagnetic invariant quantity (\ref{antisym-Fmn}). Rather, in the next subsection, we
will present a second set of gravitational invariants that yield more satisfactory results.

%%%%%%%%%%%%%%%%%%%%%%%%%%%%%%%%%%%%

\subsection{A novel set of gravitational tensors}

We will now consider a new set of gravitational quantities: two third-rank tensors
$Q_{\a\m\n}$ and $H_{\a\m\n}$ defined through the following expressions
%%%%%%%%%%%%%%%%%
\beq
Q_{\a\m\n} \equiv 2\Gamma_{\a\m\n}-\eta_{\a\n}\,\partial_\m h\,, \qquad
H_{\a\m\n} \equiv 2\Gamma_{\m\a\n}-\eta_{\a\m}\,\partial_\n h\,,
\label{QH-def}
\eeq
%%%%%%%%%%%%%%
in terms of the fully covariant form of the Christoffel symbols
$\Gamma_{\a\m\n}=\eta_{\a\rho}\,\Gamma^\rho_{\m\n}$.
Note that, although $\Gamma_{\a\m\n}$ is symmetric in the last two indices, this
symmetry is destroyed at the level of the quantities $Q_{\a\m\n}$ and $H_{\a\m\n}$
\footnote{We should also mention that the quantities $Q_{\a\m\n}$, $H_{\a\m\n}$,
$F_{\a\m\n}$ and $\G^{\a}_{\m\n}$ -- although in general are not tensors --
in the linear approximation and especially under Lorentz transformations  behave
as tensors.}.

The physical motivation for the introduction of these two gravitational tensors lies in the
fact that the following scalar combination of them
%%%%%%%%%%%%%%%
\beq
\Lambda_1 \equiv -\frac{1}{4}\,Q^{\a\m\n}\,H_{\a\m\n}\,,
\label{Lambda1-def}
\eeq
%%%%%%%%%%%%%%
can be shown, by a simple substitution of Eqs. (\ref{QH-def}), to give
exactly the Lagrangian of the gravitational field in the weak-field approximation, as this is
given in \cite{Carroll}, namely
%%%%%%%%%%%%%%%%
\beq
{\cal L}=\frac{1}{2}\,\left(\frac{1}{2}\,\partial_\m h_{\a\n}\,\partial^\m h^{\a\n} -
\partial_\m h_{\a\n}\,\partial^\a h^{\m\n} +\ \partial_\m h^{\m\n}\,\partial_\n h-
\frac{1}{2}\,\partial_\m h\,\partial^\m h\right).
\label{Lagrangian-GR}
\eeq
%%%%%%%%%%%%%%%
The variation of the above Lagrangian with respect to the metric perturbations $h_{\m\n}$
leads to the perturbed Einstein's equations (\ref{Einstein_new}). Therefore, the sheer
analogy between $\Lambda_1$ and the Lagrangian of electromagnetism, ${\cal L}_{EM}=
-F^{\m\n} F_{\m\n}/4$, makes the aforementioned scalar combination an excellent
candidate for one of the scalar invariant quantities in GEM.

In order to compute the scalar invariant quantity $\Lambda_1$, we need first the
components of the novel tensors $Q_{\a\m\n}$ and $H_{\a\m\n}$ for the different
metric ansatzes.  We will start
with the generalised form of the metric perturbations given in Eq. (\ref{case_general}):
employing the components of the $\Gamma^{\a}_{\m\n}$ quantities from 
Eq. (\ref{Christ_trad}) of the Appendix A, and the trace relation
$h=-\tilde h$ along with Eq. (\ref{trace_h}), a straightforward
calculation\,\footnote{Since the presence
of the scalar potential $\lambda$ has again no significant effect on the derived results,
for simplicity, we set again $\lambda=0$.} leads to the components of the $Q_{\a\m\n}$
and $H_{\a\m\n}$ tensors presented in Eqs. (\ref{Q_general}) and (\ref{H_general}),
respectively, in the Appendix B. Substituting these into Eq. (\ref{Lambda1-def}), we
finally find the result
%%%%%
\beq\label{Lambda1-general}
-\frac{1}{4}\,Q^{\a\m\n} H_{\a\m\n} = \frac{2}{c^4}\,(4B^2 -E^2) - R\,,
\eeq
%%%%%%%%%%%%%%%
where the sum of the sub-dominant terms $R$ has the form
%%%%%%%%%%%
\bea
\hspace*{-0.6cm} R&=& \frac{6}{c^5}\,(\partial_i \Phi)(\partial_t A^i)
+ \frac{1}{c^6}\,\left[6\,(\partial_t \Phi)^2 +
\frac{1}{2}\,(\partial_t A^i)\,(\partial_t A_i)\right] +
\frac{4}{c^7}\,\partial_t d_{ij}\,(\partial^i A^j)
\nonumber \\[1mm]
&+&\frac{1}{c^8} (\partial_j d_{ik} + \partial_k d_{ij} -\partial_i d_{jk}) (\partial^k d^{ij}
+ \partial^i d^{jk} -\partial^j d^{ik}) - \frac{1}{c^{10}}(\partial_t d_{ij})^2.
\label{R-L1-general}
\eea
%%%%%%%%%%%%%
We observe that the scalar invariant (\ref{Lambda1-general}) yields again the desired
terms $E^2$ and $B^2$ of the GEM fields -- the combination differs again from the
expected $(B^2-E^2)$, however, we note that the coefficient of 4 in front of $B^2$
matches the coefficient appearing in front of $\vec{B}$ in the corresponding expression
of the Lorentz force (\ref{Lorentz-general}), a feature that may have an underlying
significance -- in \cite{Mashhoon} it was argued that this factor is due to the spin
of the gravitational field. What is also important is the fact that
the result is completely free of any additional terms of order ${\cal O}(1/c^4)$: all
terms appearing in $R$ are suppressed by at least an additional factor of $c$ --
remarkably, for static configurations most of the additional sub-dominant terms trivially
vanish.

Let us also calculate the expression of the $\Lambda_1$ scalar invariant for the
alternative metric ansatz (\ref{case_alter}). Using the corresponding components
of $\Gamma^{\a}_{\m\n}$ presented in Eq. (\ref{Christ_alter}), and the trace $h$ 
[easily deduced from Eq. (\ref{trace_h_alter}) with $\gamma=1$], we find first the
components of the $Q_{\a\m\n}$ and $H_{\a\m\n}$ tensors; these are given in
Eqs. (\ref{Qamn_alter}) and (\ref{Hamn_alter}) of the Appendix B. Employing those in
Eq. (\ref{Lambda1-def}), we obtain the result
%%%%%
\beq\label{Lambda1-alter}
-\frac{1}{4}\,Q^{\a\m\n} H_{\a\m\n} = \frac{1}{2c^4}\,(B^2 -2E^2) -\frac{1}{4}\,\tilde R\,,
\eeq
%%%%%%%%%%%%%%%
after having used again the gauge conditions (\ref{lor6}) and (\ref{extra_v0}),
the first one for simplicity, the second as a pre-requisite for the correct form of
field equations. The sum of the sub-dominant terms $\tilde R$ now has the form
%%%%%%%%%%%
\bea
\hspace*{-0.6cm}\tilde R&=& \frac{4}{c^6}\,\left[(\partial_t \Phi)^2 +
(\partial_t A^i)\,(\partial_t A_i)\right] +
\frac{4}{c^7}\,\partial_t d_{ij}\,(\partial^i A^j)
\nonumber \\[1mm]
&+&\frac{1}{c^8} (\partial_j d_{ik} + \partial_k d_{ij} -\partial_i d_{jk}) (\partial^k d^{ij}
+ \partial^i d^{jk} -\partial^j d^{ik}) - \frac{1}{c^{10}}(\partial_t d_{ij})^2.
\label{R-L1-alter}
\eea
%%%%%%%%%%%%%
Once again, the scalar invariant (\ref{Lambda1-alter}) comes out to be free of any
additional corrections that are of the same magnitude as $E^2$ and $B^2$ -- in fact,
the next-to-leading order term of ${\cal O}(1/c^5)$ is now completely missing.
Again, for static potentials, most of the sub-dominant corrections vanish leaving
behind only an additional term suppressed by a factor of ${\cal O}(1/c^8)$. Although
the dominant combination of terms has a minimal, correct form, a superfluous factor
of 2 in front of $E^2$ destroys the perfect analogy.

Our next task is to construct a second invariant quantity in the context of GEM, the analogue
of $\tilde F\,F$ of Eq. (\ref{antisym-Fmn}). Due to the fact that the $Q$ and $H$ tensors
are third-rank tensors, there is a variety of ways that one may construct their dual
quantities by employing the antisymmetric $\epsilon^{\m\n\r\s}$ tensor. Remarkably,
almost all combinations lead to a null result for the corresponding scalar invariant
quantity. We have succeeded in finding a non-trivial result only for the following definition
of the dual form of the $Q_{\a\m\n}$ tensor
%%%%%%%%%%%%%%%%%
\beq
\tilde Q^{\a\m\n}=\epsilon^{\a\m\r\s}\,Q_{\r\s}^{\4 \n}\,.
\label{Q-dual}
\eeq
%%%%%%%%%%%%%%%%%%%%%
By using the above, one may easily construct the scalar invariant quantity
%%%%%%%%%%%%%
\beq
\Lambda_2=\tilde Q^{\a\m\n}\,H_{\a\m\n} =\epsilon^{\a\m\r\s}\,Q_{\r\s}^{\4 \n}\,H_{\a\m\n}=
Q_{\r\s\n}\,\tilde H^{\rho\s\n}\,.
\label{dualQH}
\eeq
%%%%%%%%%%%%%%
where $\tilde H^{\rho\s\n}=\epsilon^{\rho\s\a\m}\,H_{\a\m}^{\4 \n}$.
In addition, two more scalar invariant quantities could be constructed, namely
$\Lambda_3=\tilde Q^{\a\m\n}\,Q_{\a\m\n}$ and  $\Lambda_4=\tilde H^{\a\m\n}\,H_{\a\m\n}$.
In fact, one may easily show all three scalar invariants $\Lambda_2$, $\Lambda_3$ and
$\Lambda_4$ are identical: using the definition of the dual tensor in each case
and the expressions (\ref{QH-def}) of the $Q$ and $H$ tensors, one arrives at the
general relation
%%%%%%%%%%%%%
\beq
\tilde Q^{\a\m\n}\,H_{\a\m\n} =
-\tilde Q^{\a\m\n}\,Q_{\a\m\n} = -\tilde H^{\a\m\n}\,H_{\a\m\n} =
4 \epsilon^{\a\m\rho\s}\Gamma_{\rho\s}^{\4 \n} \Gamma_{\m\a\n}\,.
\label{scalar-relation}
\eeq
%%%%%%%%%%%%%%%%%%
The above result eliminates the apparent freedom in the construction of the second
invariant quantity, and allows us to choose any of the above combinations as its
functional form. In practice, its expression will be calculated by using the
components of the Christoffel symbols. 

We will calculate the above quantity for both metric perturbations ansatzes
(\ref{case_general}) and (\ref{case_alter}). Starting with the generalised
traditional ansatz for the metric perturbations (\ref{case_general}) and
using the corresponding components of $\Gamma^{\a}_{\m\n}$ from
Eq. (\ref{Christ_trad}), we obtain the following result
%%%%%%%%%%%%%%%%%%%
\beq
\frac{1}{16}\,\tilde Q^{\a\m\n}\,H_{\a\m\n}=-\frac{8}{c^4}\,\vec{E} \cdot \vec{B} + Y,
\label{L2_general}
\eeq
%%%%%%%%%%%%%%%%%%
where the quantity $Y$ stands for the sum of the sub-dominant terms, namely
%%%%%%%%%%%%%
\beq
Y=-\frac{1}{c^5}\,\vec{B}\cdot \partial_t \vec A -
\frac{1}{c^6}\,e^{0ilk}\,\partial_k d_l^{\,\, j}
\left(\partial_i A_j + \partial_j A_i -\frac{2}{c^3}\,\partial_t d_{ij}\right).
\label{Y_general}
\eeq
%%%%%%%%%%%%
The quantity $\tilde Q^{\a\m\n} H_{\a\m\n}$ is found to have indeed the desired form
being proportional to the internal product of the GEM fields $\vec{E}$ and $\vec{B}$,
as in Eq. (\ref{antisym-Fmn}). The additional terms appearing in $Y$ are sub-dominant
- again, for static configurations
the only surviving term is of the order of ${\cal O}(1/c^6)$. In addition, due to the
way the GEM fields combine in this invariant, any superfluous numerical factor appears
as an overall multiplicative factor in front of their internal product, and may be easily
absorbed into the definition of the scalar quantity, as was indeed done in Eq. (\ref{L2_general}).

Moving finally to the alternative ansatz for the metric perturbations
(\ref{case_alter}), and employing the $\Gamma^{\a}_{\m\n}$ components -- displayed in
Eq. (\ref{Christ_alter}) -- in the $\Lambda_2$ scalar
invariant quantity, we find the result
%%%%%%%%%%%%%%%%%%%
\beq
\frac{1}{4}\,\tilde Q^{\a\m\n}\,H_{\a\m\n}=-\frac{5}{c^4}\,\vec{E} \cdot \vec{B} + Y,
\label{L2_alter}
\eeq
%%%%%%%%%%%%%%%%%%
where the quantity $Y$ now stands for the following combination
%%%%%%%%%%%%%
\beq
Y=-\frac{2}{c^5}\,\vec{B}\cdot \partial_t \vec A -
\frac{1}{c^6}\,e^{0ilk}\,\partial_l d_k^{\,\, j}
\left(\partial_i A_j + \partial_j A_i -\frac{2}{c^3}\,\partial_t d_{ij}\right).
\label{Y_alter}
\eeq
%%%%%%%%%%%%
Once again, the $\Lambda_2$ invariant yields the correct dependence on the GEM
fields, while any additional terms are sub-dominant. We observe that the
use of the two different metric ansatzes has resulted in very similar results
for the dual scalar invariant quantity, namely in 
Eqs. (\ref{L2_general})-(\ref{Y_general}) and (\ref{L2_alter})-(\ref{Y_alter});
differences appear only in the numerical factors and not in the functional
form, a feature that may hint to some kind of universality of this invariant
quantity for a class of metric ansatzes in the context of GEM.

%%%%%%%%%%%%%%%%%%%%%%%%%%%%%%%%%%%%%%%%%%%%%%%%%%%%%%%%%%%%%%%%%%%%%%%

\section{Coordinate transformation and GEM gauge invariance}
\setcounter{equation}{0}

Let us consider the following coordinate transformation 
%%%%%%%%%%%%
\beq
x^\m \rightarrow x'^\m=x^\m -\epsilon^\m(x^\rho)\,,
\label{coord-trans}
\eeq
%%%%%%%%%%%%%%%%%%%
where $\epsilon^\mu$ is an arbitrary vector that, in general, depends on all
spacetime coordinates. If we want the above coordinate transformation to be
a diffeomorphism, then, under the decomposition (\ref{metric}) for the metric
tensor and the assumption that $\epsilon^\mu$ is also a small quantity,
we obtain the perturbations transformation \cite{Landau}\cite{Carroll}
%%%%%%%%%%%%%%%%%%%
\beq
h'_{\m\n}=h_{\m\n} + \epsilon_{\m,\n} + \epsilon_{\n,\m}\,.
\label{pert-trans}
\eeq
%%%%%%%%%%%%%%%%%
One may verify that the perturbed Einstein's equations
remain invariant under the above transformation of $h_{\m\n}$, a result that
is known as the gauge invariance of the linearised gravitational theory~\footnote{Here,
we work at the lowest order and thus the energy-momentum tensor $T_{\m\n}$ is
taken to be independent of $h_{\m\n}$.}.

Using the above transformation relation for $h_{\m\n}$, we may find that
the transformation rule for the trace is: $h'=h+2 \epsilon^\r_{\2,\r}$. 
Employing this, the transformation relation for the new perturbations $\tilde h_{\m\n}$
takes the form
%%%%%%%%%%%%
\beq
\tilde h'_{\m\n} =\tilde h_{\m\n} + \epsilon_{\m,\n} + \epsilon_{\n,\m}
-\eta_{\m\n}\,\epsilon^\r_{\2,\r}\,.
\label{pert-new-trans}
\eeq
%%%%%%%%%%%%
One may easily check that under the above, the novel form of the
perturbed Einstein's equations (\ref{Einstein_new}) remains indeed
invariant.

As the components of the perturbations $\tilde h_{\m\n}$ are directly
related to the scalar $\Phi$ and vector potential $\vec{A}$, any coordinate
transformation causes a change in the potentials themselves. In this section,
we would like to investigate whether the gauge invariance of the gravitational
linearised theory amounts to a gauge invariance of the GEM fields $\vec{E}$
and $\vec{B}$. This question has been addressed in the literature before
\cite{Mashhoon} but for a more special choice of the metric perturbations
and under some simplifying assumptions. Here, we will employ the more 
successful ansatzes (\ref{case_general}) and
(\ref{case_alter}) -- from the point of view of the restoration of the form of
both Maxwell's and Lorentz equations -- and attempt to find the most general configuration
for the arbitrary vector $\epsilon^\m(x^\rho)$ for which the gauge invariance
of the GEM fields holdσ.

We will start with the generalised traditional ansatz (\ref{case_general}).
The transformation rule (\ref{pert-new-trans}) for the $\tilde h_{00}$ and
$\tilde h_{0i}$ components lead to the following transformation of the 
potentials $\Phi$ and $\vec{A}$
%%%%%%%%%%%%%%%%%
\bea
\Phi &\rightarrow & \Phi' =\Phi  +\frac{c^2}{4}\,(\partial_0 \epsilon^0
-\partial_i \epsilon^i)\,, \label{Phi-trans}\\[1mm]
A_i &\rightarrow & A'_i =A_i  +\frac{c^2}{2}\,(\partial_i \epsilon_0+
\partial_0 \epsilon_i)\,.
\label{A-trans}
\eea
%%%%%%%%%%%%%%%%
The transformation rule of the $\tilde h_{ij}$ components contains the 
changes of both the scalar potential $\lambda$ and the symmetric tensor
$d_{ij}$. In order to disentangle these two transformation relations, we
employ also the transformation of the trace $h$
and the relation $\tilde h= -h$. Then, we find
%%%%%%%%%%%%%%
%%%%%%%%%%%%%%%%%
\bea
\lambda &\rightarrow & \lambda' =\lambda  -\frac{c^4}{2}\,(\partial_0 \epsilon^0
+\frac{1}{3}\,\partial_i \epsilon^i)\,, \label{lambda-trans}\\[1mm]
d_{ij} &\rightarrow & d'_{ij} =d_{ij}  +\frac{c^4}{2}\,(\partial_i \epsilon_j +
\partial_j \epsilon_i -\frac{2}{3}\,\eta_{ij}\,\partial_k \epsilon^k)\,.
\label{d-trans}
\eea
%%%%%%%%%%%%%%%%
Although the GEM fields depend only on $\Phi$ and $\vec{A}$, the presence of
$\lambda$ and $d_{ij}$, along with their changes under the aforementioned
transformation rules, is imperative for the coordinate transformation to ``close''
-- we will return to this point later. 

Let us now examine how the GEM fields change under the corresponding changes
of the potentials. Their transformation rules are found to be~\footnote{Note that,
in the case of a general coordinate transformation,
we should have taken into account how the derivatives $\partial_\m$ change as
well, i.e. $\partial_\m \rightarrow \partial'_\m = \partial_\m + 
\epsilon^\r_{\2,\m}\,\partial_\r$. However, the latter term when acting on the
perturbations $\tilde h_{\m\n}$ is of quadratic order and thus, in the linear
approximation, it is ignored.}
%%%%%%%%%%%%%%%%
\bea
E^i \,\rightarrow\, E'^i &=& 
E^i + \frac{c^2}{4}\,(\partial^2_0 \epsilon_i +\partial_i \partial_k \epsilon^k)\,, 
\nonumber \\[2mm]
B^k \,\rightarrow\, B'^k &=& B^k +\frac{c^2}{4}\,\epsilon^{kij}\,
\partial_0\partial_i \epsilon_j\,.
\label{EB-trans}
\eea
%%%%%%%%%%%%%%
In the context of Electromagnetism, the gauge invariance is the invariance
of the electric $\vec{E}$ and magnetic field $\vec{B}$ under the following
changes of the potentials
%%%%%%%%%%%%%
\beq
\Phi \rightarrow \Phi' =\Phi -\partial_0 \Lambda\,,
\qquad A^i \rightarrow A'^i =A^i + \partial_i \Lambda\,,
\label{EM-gauge}
\eeq
%%%%%%%%%%%%%%%%%
where $\Lambda$ is an arbitrary scalar function. In the context of GEM, and
for the traditional ansatz (\ref{case1}), a similar type of gauge invariance was
discussed in \cite{Mashhoon}: in there, it was assumed that $\epsilon^i=0$ and
only the temporal component $\epsilon^0$ was kept; then, the functional forms
of the Eqs. (\ref{Phi-trans})-(\ref{A-trans}) and (\ref{EM-gauge}) exactly match
and the gauge invariance indeed holds.

However, as with the case of the additional components of the Einstein's
equations -- which are not necessary for the analogy with the EM but are 
nevertheless present -- also here, the presence of the spatial components
$\epsilon^i$ is just another manifestation of the broader
structure of the General Theory of Relativity compared to the U(1) gauge field
theory of EM. Over-restrictive choices for the ``additional'' components, that
we do not seem to need, leads either to unphysical constraints in the theory 
-- if they are not properly taken into account, as discussed in Section 3 --
or they unnecessarily restrict the space of a symmetry in the theory as with
the components of $\epsilon^\m$. In fact, an exact matching of
Eqs. (\ref{Phi-trans})-(\ref{A-trans}) with the transformations (\ref{EM-gauge})
can still be achieved, even for a non-trivial $\epsilon^i$, provided that the 
relations $\partial_0 \epsilon_i=\partial_i \epsilon^i=0$ hold: then, the gauge
invariance of the fields is valid with the arbitrary scalar function given by the
relation $\Lambda=-c^2 \epsilon_0/4$.

However, there is an even broader class of transformations for the potentials
that leave the GEM fields (\ref{EB-trans}) invariant. As their
expressions show, we merely need to satisfy the following constraints
%%%%%%%%%%%%%%
\bea
\partial^2_0 \epsilon_i +\partial_i \partial_k \epsilon^k &=& 0\,, 
\label{cond2a-epsilon}\\[1mm]
\partial_0\partial_i \epsilon_j &=&0\,.
\label{cond2b-epsilon}
\eea
%%%%%%%%%%%%%
To the above, we will add an additional constraint that follows from the demand
that the transverse gauge condition - as it helps to restore the form of the
Maxwell's equations -- holds in all coordinate systems, i.e.
$\partial_\n\tilde h^{\m\n}=\partial_\n\tilde h'^{\m\n}=0$; this has been
broadly assumed in the literature including the traditional case, and leads to
the relation
%%%%%%%%%%%%%%%%
\beq
\partial^2 \epsilon^\m=(\partial_0^2- \partial_i^2)\,\epsilon^\m=0\,.
\label{cond3-epsilon}
\eeq
%%%%%%%%%%%%%%%%%%%%%
The temporal component $\epsilon^0$ satisfies only Eq. (\ref{cond3-epsilon}),
and thus will have the general form $\epsilon^0=\epsilon^0 (\vec{x}-c t)$.
On the other hand, the spatial components $\epsilon^i$ need to satisfy all
three constraints (\ref{cond2a-epsilon})-(\ref{cond3-epsilon}). Equation
(\ref{cond2b-epsilon}) demands that the time
and space dependence should be separated, namely $\epsilon^i=f_i(\vec{x})+
g_i(t)$, for $i=1,2,3$. Then, the spatial components of Eq. (\ref{cond3-epsilon})
dictate that this time and space dependence should be at most quadratic.
In any case, we conclude that the space of the gauge symmetry of the theory
can be significantly enlarged if we avoid the over-restrictive assumption of
vanishing $\epsilon^i$.

The above analysis can be repeated for the alternative ansatz of the metric
perturbations (\ref{case_alter}). The transformation rules for $\Phi$ and $\vec{A}$
are found to be given again by Eqs. (\ref{Phi-trans})-(\ref{A-trans}) under the
redefinitions $\Phi \rightarrow \Phi/4$ and $\vec{A} \rightarrow \vec{A}/2$.
However,
in the next step we encounter a problem: considering the transformation
rule for the trace $\tilde h$, we obtain two different results, one if
we use its transformation rule $\tilde h'=\tilde h -2 \epsilon^\r_{\2,\r}$
and one by employing its explicit form in terms of $\Phi$, Eq. (\ref{trace_h_alter}),
and use the transformation rule for $\Phi$ itself. This is a sign that the set
of transformation rules does not ``close" as it is. The obstacle is easily
overcome by adding to the $\tilde h_{ij}$ component a term proportional to
a scalar potential $\lambda$ and suppressed by $1/c^4$. Then, the
$\tilde h_{ij}$ component reads
%%%%%%%%%%%%%%%%%%
\beq
\tilde h_{ij} = \left(-\frac{\Phi}{c^2}+\frac{\lambda}{c^4}\right) \eta_{ij}
+\frac{1}{c^4}\,d_{ij}\,.
\label{hij-new}
\eeq
%%%%%%%%%%%%%%%%%
As in the case of the generalised traditional ansatz (\ref{case_general}), this term
plays an insignificant role, and its presence causes no modification to the conclusions
of the previous sections regarding the alternative ansatz. However, its presence
will help the set of transformation rules to ``close": indeed, assuming the form
(\ref{hij-new}) for the $\tilde h_{ij}$ component, we arrive with no inconsistencies
at the complementary transformation rule
%%%%%%%%%%%%%%%%
\beq
\lambda \rightarrow \lambda' =\lambda - \frac{4c^4}{3}\,\partial_i \epsilon^i\,,
\label{lambda-alter-trans}
\eeq
%%%%%%%%%%%%%%%
while the transformation rule for $d_{ij}$ is given again by Eq. (\ref{d-trans})
under the change $d_{ij} \rightarrow d_{ij}/2$.

The corresponding changes to the GEM fields $\vec{E}$ and $\vec{B}$ are given
by Eqs. (\ref{EB-trans}), with the
only difference being the rescalings $\vec{E} \rightarrow \vec{E}/4$ and
$\vec{B} \rightarrow \vec{B}/4$. As a result, the most general configuration
for the arbitrary vector $\epsilon^\m$ that ensures the gauge invariance
of the GEM fields are still given by the set of Eqs.
(\ref{cond2a-epsilon})-(\ref{cond3-epsilon}), and the same results hold. 
Once again, there is no need to assume a vanishing $\epsilon^i$ and the
space of the symmetry is enlarged compared to previous
treatments~\footnote{Note that, according to Eq. (\ref{lambda-alter-trans}),
the absence of $\lambda$ would be admissible, and no inconsistencies with the
closure of the transformation rules would arise, if the condition 
$\partial_i \epsilon^i=0$ was demanded. 
However, as we saw earlier this leads to a more restrictive form of the
vector $\epsilon^\m$ than it is necessary. For this reason, above, we have
followed instead the option of the introduction of the $\lambda$ term that, while
not affecting any of our results, helps to keep the space of the symmetry
as large as possible.}.

The inconsistency that has arisen in applying the transformation rule
(\ref{pert-new-trans}) to the case of the alternative ansatz is, we believe,
a generic one that appears when an over-simplifying assumption is made for the
form of the metric perturbations. Although the problem for the alternative
ansatz was fixed quite easily, a more serious consistency problem will
be present for ansatzes that are ``built" to be very simple. For example,
one may easily see that the traditional ansatz (\ref{case1}), where the
$\tilde h_{ij}$ components were altogether ignored, faces a similar problem;
unless one restores the spatial components of the metric perturbations -- an
approach that we have consistently followed in this work for a number of
additional reasons -- the problem cannot be overcome.

%%%%%%%%%%%%%%%%%%%%%%%%%%%%%%%%%%%%%%%%%%%%%%%%%%%%%%%%%%%%%%%%%%%%%%%%%%

\section{Discussion and Conclusions}

The striking similarity between the gravitational and electromagnetic forces at
classical level, that was found to hold also in the context of the General Theory
of Relativity, has attracted a lot of attention over a century-long period.
The framework of Gravito-electromagnetism, the theory that describes the dynamics
of the gravitational field in terms of quantities met in electromagnetism, has 
also been the area of an intense activity over the years and, provided a new
perspective on the description and understanding of the gravitational field. 

In this work, we have adopted the linear-approximation approach where the 
gravitational perturbations $h_{\m\n}$ involve directly the GEM potentials
$\Phi$ and $\vec{A}$. The perturbed Einstein's field equations then reduce to
a set of Maxwell-like equations for the GEM fields $\vec{E}$ and $\vec{B}$.
In this, and in a previous work of ours \cite{BK1}, we have shown that the
form of the gravitational perturbations is of paramount importance for the
successful analogy between gravity and EM. In section 2, we reviewed the
weak points of the analysis based on the traditional ansatz (\ref{case1}) for the
gravitational perturbations \cite{Mashhoon}: (i) the Maxwell's equations are
restored only for a static vector potential $\vec{A}$, (ii) the Lorentz
equation is reproduced but, again, only for a static scalar potential $\Phi$,
(iii) the additional components of the field equations and gauge condition
can impose unphysical constraints to the fields or matter
distribution of the theory.

In the light of the above results, in Section 3, we introduced two novel forms
of metric perturbations. The first one, Eq. (\ref{case_general}), was a generalisation
of the traditional ansatz and allowed small but non-vanishing spatial components
$\tilde h_{ij}$ of the metric perturbations. Although these are not involved in the
derivation of Maxwell's equations, we demonstrated that they have important
implications for the analysis: in their absence, the additional component of the
transverse gauge condition {\it demands} the staticity of the vector potential
$\vec{A}$; in its presence, all terms in Maxwell's equations, involving 
time-derivatives of $\vec{A}$, are restored, and all additional constraints are
rendered harmless, simply connecting the form of $\tilde h_{ij}$
to the distribution of matter in the theory.

The expression of the Lorentz equation, though, still suffered from the presence
of additional terms. In search of an ansatz that could perhaps achieve both
objectives at once, i.e. restore Maxwell's equations and the Lorentz equation, in
Section 3.2 we proposed the so-called alternative ansatz (\ref{case_alter}). This
comprised a variant of a similar ansatz for the metric perturbations
proposed in our previous work \cite{BK1}, its core idea being the introduction of
the scalar potential $\Phi$ into the spatial components $\tilde h_{ij}$, too.
In the context of the present work, that ansatz was extended and showed
to lead to the correct, minimal form of the Lorentz equation. We also demonstrated
that its use avoids again any unphysical constraints in the theory,
and restores Maxwell's equations apart from a superfluous coefficient
of 4, that could be perhaps absorbed into the redefinition of the matter/charge density
in the context of GEM. 

Pursuing further the analogy between gravity and EM, in Section 4 we searched for
scalar invariant quantities defined in terms of gravitational tensors in the 
context of the linear approximation. In Section 4.1, we defined a novel 3rd-rank
tensor $F_{\a\m\n}$ -- an analogue of the electromagnetic field strength tensor
$F_{\m\n}$ -- and the scalar combination $F^{\a\m\n} F_{\a\m\n}$ was computed.
For both metric ansatzes (\ref{case_general}) and (\ref{case_alter}), this combination 
resembled the one of $F^{\m\n} F_{\m\n}$ in EM but contained
additional terms of the same order as $\vec{E}^2$ and $\vec{B}^2$. In
Section 4.2, we proceeded to define two novel 3rd-rank
tensors $Q^{\a\m\n}$ and $H^{\a\m\n}$ motivated by the fact that the combination
$-Q^{\a\m\n}\,H_{\a\m\n}/4$ matches the gravitational Lagrangian in the linear
approximation. Here, the results were very encouraging: the aforementioned
scalar quantity was indeed proportional to a combination of $\vec{E}^2$
and $\vec{B}^2$ containing no additional terms of the same order.
Enforced by these results, we then defined a second scalar
invariant quantity of the form $\tilde Q^{\a\m\n} H_{\a\m\n}$, where
$\tilde Q^{\a\m\n}$ is the dual of $Q^{\a\m\n}$ -- an analogue
again of the $\tilde F^{\m\n} F_{\m\n}$ quantity in EM -- that was found to be
proportional to the internal product $\vec{E} \cdot \vec{B}$, with no
additional corrections of the same order.

Let us note at this point that the above 3rd-rank tensors were defined in
a covariant way, and thus may be used to build scalar quantities for arbitrary
gravitational backgrounds in the context of the linearised theory. Their
tensorial character is, strictly speaking, valid only in the context of the
linear theory, however, the motivation for their introduction has a solid,
physical base: they may be used to re-write, in a covariant way, either the
field equations or the Lagrangian of the theory. In the absence of the physical
connection between the exact and linear approach, we may not be certain about
the role that these tensors may have in the context of an exact description.
However, since an exact analysis should have a smooth limit to the weak-field
description, and thus to the results found here, our proposed tensors are
bound to be present in the context of the exact theory, possibly as part of
a more fundamental tensorial quantity.

In the last section, section 5, we investigated in detail the conditions under
which the gauge invariance of the linearised gravitational theory leads to
a gauge invariance of the GEM fields $\vec{E}$ and $\vec{B}$. For a small
displacement of the coordinates, parametrised by an arbitrary vector $\epsilon^\m$,
the transformation rule for the metric perturbations $h_{\m\n}$ was found,
and from that the transformation relations for the GEM potentials and fields
were derived. As has been shown in the literature \cite{Mashhoon}, the latter
transformation relations reduce to the usual gauge transformations of EM
under the assumption that the spatial components of $\epsilon^\m$ vanish.
Here, using our two novel ansatzes for the metric perturbations
(\ref{case_general}) and (\ref{case_alter}), 
we have demonstrated that a gauge invariance of the GEM fields $\vec{E}$ and
$\vec{B}$ holds for a much more general form of the arbitrary vector
$\epsilon^\m$. The conditions that the temporal and spatial components of
$\epsilon^\m$ need to satisfy were derived, and their general configuration
was determined. In this way, an over-restrictive assumption regarding the form of
the arbitrary vector $\epsilon^\m$ was avoided, and the space of the symmetry
present in the theory was enlarged compared to previous studies.

In the previous work of ours \cite{BK1}, we had also posed the question whether
a tensorial theory based on the linearised Einstein's equations, but with the
constant $k$ being different from its usual value of $8\pi G/c^4$, could exactly
re-produce the theory of Electromagnetism. Had we repeated that analysis,
a number of satisfactory results would have emerged: (i) for the generalised
traditional ansatz (\ref{case_general}),
all Maxwell's EM equations would have been exactly reproduced under the assumption
that $k=-2\pi/c^4$; however, the additional terms in the expression of the
Lorentz force would now have dire phenomenological consequences, (ii) for
the alternative ansatz (\ref{case_alter}), though, the expression of the Lorentz force
would exactly match the one in EM with the only additional corrections being of order
${\cal O}(1/c^4)$ and thus negligible even in the non-relativistic limit; 
in addition, the set of Maxwell's equations would have been again reproduced,
now under the assumption that $k=-8\pi/c^4$. We thus conclude that the use
of the alternative ansatz has finally provided the proper form of metric perturbations
that, in conjunction with Einstein's tensorial equations, has reproduced the complete
mathematical framework of the theory of electromagnetism.

Naturally, the above suggestion does not aim at replacing the theory of electromagnetism
by a tensorial theory; it rather aims at investigating how far the analogy between
gravity and electromagnetism extends, given the persistent difficulty in finding a
common mathematical framework to unify gravity and gauge forces. Moreover, as it
has been the objective of the theory of GEM itself, this analogy may open new
directions and lead to the discovery of new phenomena in gravitational physics
guided by phenomena already observed in EM. 
Although the formulation of an exact description of the analogy between
gravity and EM is the ultimate objective, that exact description owes to 
have a clearly defined limit in the weak-field approximation, that is the
natural framework for the study of gravitational phenomena for a distant
observer. In addition, this exact description is bound to face all the problems
of the linear theory in case the latter employs the wrong assumptions.
Therefore, in this work we have attempted to provide a remedy for a number
of weak points that the traditional GEM analysis suffers from.
We believe that the two novel ansatzes that we have proposed here have come
a long way in overcoming previous inconsistencies of GEM and, in conjunction
with the proposed 3-rank tensors, in achieving the best analogy between gravity
and EM in the literature so far. Our results have also improved a previous
analysis of ours in the sense that the set of perturbed equations derived in
this work are free of any unphysical constraints and valid both far away and
close to the sources of the gravitational field; this, may prove to be of
significant importance in future applications of our formalism to phenomena
in gravitational physics, especially in the light of the recent discovery of
gravitational waves whose study is based on the linear approximation method.

%%%%%%%%%%%%%%%%%%%%%%%%%%%%%%%%%%%%%%%%%%%%%%%%%%%%%%%%%%%%%%%%%%%%%%%%%%%%%%%%

%\begin{acknowledgements}
%Part of this work was supported by the COST Action MP1210 ``The String Theory Universe''.
%\end{acknowledgements}

\begin{appendix}

\section{Components of the Christoffel symbols}
\setcounter{equation}{0}

Here, we present the components of the Christoffel symbols for both 
metric perturbations ansatzes that are necessary for the derivation of the
Lorentz equation in each case.

Employing the expression of the Christoffel symbols (\ref{Christoffel}) in
the linear approximation and of the initial perturbations $h_{\mu\nu}$ from
the line-element (\ref{line-element-general}) for the case of the generalised
traditional ansatz, we find the following components
%%%%%%%%%%%%%%
\bea
\G^i_{00} &=& 
%\frac{1}{2}\,\h^{ij}\left(\frac{2}{c}\, \aa_t h_{0j}-\aa_j h_{00}\right)=
\frac{1}{c^2}\,\partial_i \Phi-
\frac{3}{2c^4}\,\partial_i \lambda +\frac{2}{c^3}\,\partial_t A^i\,, \quad
%\nonumber \\[1mm]
%%%%
\G^i_{0j} =%&=& 
%\frac{1}{2}\,\h^{ik}\left(\an[j]h_{0k}-\an[k]h_{0j}\right)=
\frac{1}{c^2}\,F_{ij} - \frac{1}{c^3}\,\delta^i_j \left(\partial_t \Phi
+\frac{1}{2c^2}\,\partial_t \lambda \right)
+ \frac{1}{c^5}\,\partial_t d^i_{\,\, j}\,,  \nonumber \\[1mm]
%%%%
\G^i_{kj} &=& 
%\frac{1}{2}\,\h^{il}\left(h_{lj,k}+h_{lk,j}-h_{jk,l}\right)=
-\frac{1}{c^2}\left(\delta^i_j \,\partial_k \Phi +
\delta^i_k \,\partial_j \Phi -\delta_{kj} \,\partial_i \Phi \right) \nonumber \\[1mm]
&-& \frac{1}{2c^4}\left(\delta^i_j \,\partial_k \lambda +
\delta^i_k \,\partial_j \lambda -\delta_{kj} \,\partial_i \lambda \right)+
 \frac{1}{c^4}\left(\partial_k d^i_{\,\, j} + \partial_j d^i_{\,\, k}-
\partial^i d_{kj} \right). \label{Christ_trad}
\eea
%%%%%%%%%%%%%%%
Note that above we have used the definition 
$F_{ij} \equiv \partial_i A_j -\partial_j A_i$.

In the case of the alternative ansatz for the metric perturbations, the
original perturbations $h_{\mu\nu}$ may be deduced from the line-element
(\ref{line-element-alter}). Then, the Christoffel symbols  are found to be
%%%%%%%%%%%%%%
\bea
\G^i_{00} &=& 
%\frac{1}{2}\,\h^{ij}\left(\frac{2}{c}\, \aa_t h_{0j}-\aa_j h_{00}\right)=
\frac{(1+3\gamma)}{4c^2}\,\partial_i \Phi +\frac{1}{c^3}\,\partial_t A^i\,, \qquad
%%%%
\G^i_{0j} = 
%\frac{1}{2}\,\h^{ik}\left(\an[j]h_{0k}-\an[k]h_{0j}\right)=
\frac{1}{2c^2}\,F_{ij} - \frac{(1-3\gamma)}{4c^3}\,\delta^i_j \,\partial_t \Phi
+ \frac{1}{2c}\,\partial_t \tilde h^i_{\,\, j}\,, \nonumber
\label{Christ_alter} \\[1mm]
%%%%
\G^i_{kj} &=& 
%\frac{1}{2}\,\h^{il}\left(h_{lj,k}+h_{lk,j}-h_{jk,l}\right)=
-\frac{(1-3\gamma)}{4c^2}\left(\delta^i_j \,\partial_k \Phi +
\delta^i_k \,\partial_j \Phi -\delta_{kj} \,\partial_i \Phi \right)
+ \frac{1}{2}\left(\partial_k \tilde h^i_{\,\, j} + \partial_j \tilde h^i_{\,\, k}-
\partial^i \tilde h_{kj} \right). 
\eea
%%%%%%%%%%%%%%%

\section{Components of the novel gravitational tensors}
\setcounter{equation}{0}

By using the components of the perturbations $\tilde h_{\m\nu}$, as these
appear in the generalised traditional ansatz (\ref{case_general}), we find the following
explicit forms for the components of the new tensor $F_{\alpha\m\n}$,
defined in Eq. (\ref{Famn}):
%%%%%%
\begin{eqnarray}
&& F_{000}=-\frac{2}{c^2}\,\aa_i A^i, \,\,
F_{i00}=\frac{4 E^i}{c^2} - \frac{2}{c^4}  \left(\aa_i \l + \aa^j d_{ij} \right), \,\,
F_{ij0}=- \frac{2}{c^2} F_{ij} + \frac{2}{c^5}\aa_t \left(\l\,\h_{ij} +d_{ij} \right), 
\label{Famn_general} \nonumber \\[2mm]
&&F_{00i}=\frac{4}{c^2} \aa_i \F, \,\,\,
F_{0ij}= \frac{2}{c^2} \left[(\aa_i A_j + \aa_j A_i) -
\h_{ij} \left(\frac{2}{c}\,\aa_t \F + \aa_k A^k \right) \right] - 
\frac{2}{c^5} \aa_t (\l\, \h_{ij} +d_{ij}), \nonumber \\[2mm]
&&F_{ijk} = \frac{2}{c^4} \bigl[ \aa_j (\l \h_{ik} + d_{ik}) - (j \leftrightarrow k) -
(j \leftrightarrow i) \bigr]
-\frac{2 \h_{jk}}{c^3} \left[ \aa_t A_i +\frac{1}{c} \,\aa^l (\l \h_{il} + d_{il}) \right]. 
\end{eqnarray}
%%%%%%

Similarly, employing the components of $\tilde h_{\m\n}$ of the alternative
ansatz (\ref{case_alter}), we find the results
%%%%%%
\begin{eqnarray}
F_{000} &=& -\frac{1}{c^2} \aa_i A^i, \qquad F_{i00}= \frac{1}{c^3} \partial_t A_i -
\frac{1}{c^4} \partial^j d_{ij}, \qquad  F_{00i} =\frac{1}{c^2}\aa_i \F, \nonumber \\[2mm]
F_{ij0}&=& -\frac{F_{ij}}{c^2} - \frac{\aa_t \Phi}{c^3} \h_{ij} +\frac{1}{c^5} \partial_t d_{ij},
\quad
F_{0ij} = \frac{1}{c^2} \left[( \aa_iA_j +\aa_jA_i) -  \h_{ij} \aa_k A^k\right] 
-\frac{\partial_t d_{ij}}{c^5},   \nonumber \\[2mm]
F_{ijk} &=& -\frac{1}{c^2} \left(\h_{ij}\,\aa_k \F +\h_{ik}\,\aa_j \F -2\h_{jk}\,\aa_i \F \right)
- \frac{1}{c^3}\,\h_{jk}\,\aa_t A_i -\frac{1}{c^4}\,\eta_{jk}\,\partial^l d_{il} \nonumber \\[2mm] 
&+& \frac{1}{c^4}\,(\partial_j d_{ik} + \partial_k d_{ij} -\partial_i d_{jk})\,. 
\label{Famn_alter}
\end{eqnarray}
%%%%%%%%%%%%%%%%

We now move to the components of the gravitational tensors $Q_{\a\m\n}$ and
$H_{\a\m\n}$ defined in Eqs. (\ref{QH-def}). For the generalised ansatz (\ref{case_general})
(where we have set for simplicity $\lambda=0$),
we find the following components of the $Q_{\a\m\n}$ 
%%%%%%%%%%%%%%%%%
\begin{eqnarray}
Q_{000} &=& \frac{6}{c^3}\,\aa_t \Phi, \qquad Q_{i00}= \frac{4}{c^3}\,\partial_t A_i -
\frac{2}{c^2}\,\partial_i \Phi, \qquad  Q_{00i} =\frac{2}{c^2}\,\aa_i \F, \nonumber \\[2mm]
Q_{0i0} & =& \frac{6}{c^2}\,\aa_i \F, \qquad Q_{ij0} = -\frac{2}{c^2}F_{ij} - 
\frac{2}{c^3}\,\aa_t \Phi\,\h_{ij} +\frac{2}{c^5}\,\partial_t d_{ij},
\label{Q_general}  \\[2mm]
Q_{0ij} &=& \frac{2}{c^2} ( \aa_iA_j +\aa_jA_i) + \frac{2 \aa_t \Phi}{c^3} \h_{ij} 
-\frac{2 \partial_t d_{ij}}{c^5},  \,\,\,
Q_{i0j} = -\frac{2 F_{ij}}{c^2} +
\frac{2 \aa_t \Phi\,\h_{ij}}{c^3} +\frac{2 \partial_t d_{ij}}{c^5},
\nonumber \\[2mm]
Q_{ijk} &=& -\frac{2}{c^2} \left(\h_{ij}\,\aa_k \F -\h_{ik}\,\aa_j \F -\h_{jk}\,\aa_i \F \right)
+ \frac{2}{c^4}\,(\partial_j d_{ik} + \partial_k d_{ij} -\partial_i d_{jk})\,. \nonumber
\end{eqnarray}
%%%%%%%%%%%%%%%%
and $H_{\a\m\n}$ tensors
%%%%%%%%%%%%%%%%%
\begin{eqnarray}
H_{000} &=& \frac{6}{c^3}\,\aa_t \Phi, \qquad H_{i00}=
\frac{2}{c^2}\,\partial_i \Phi, \qquad  H_{00i} =\frac{6}{c^2}\,\aa_i \F, \nonumber \\[2mm]
H_{0i0} & =& \frac{4}{c^3}\,\partial_t A_i -\frac{2}{c^2}\,\aa_i \F, \qquad 
H_{ij0} = \frac{2}{c^2}F_{ij} +
\frac{2}{c^3}\,\aa_t \Phi\,\h_{ij} +\frac{2}{c^5}\,\partial_t d_{ij},
\label{H_general}   \\[2mm]
H_{0ij} &=&-\frac{2 F_{ij}}{c^2} -
\frac{2 \aa_t \Phi}{c^3} \h_{ij} +\frac{2 \partial_t d_{ij}}{c^5}, 
\,\,\,
H_{i0j} =  \frac{2}{c^2} ( \aa_iA_j +\aa_jA_i) + \frac{2 \aa_t \Phi }{c^3} \h_{ij}
-\frac{2 \partial_t d_{ij}}{c^5}
\nonumber \\[2mm]
H_{ijk} &=& \frac{2}{c^2} \left(\h_{ij}\,\aa_k \F +\h_{ik}\,\aa_j \F -\h_{jk}\,\aa_i \F \right)
+ \frac{2}{c^4}\,(\partial_j d_{ik} + \partial_k d_{ij} -\partial_i d_{jk})\,. \nonumber
\end{eqnarray}
%%%%%%%%%%%%%%%% 

We follow a similar analysis for the case of the alternative ansatz (\ref{case_alter})
(with $\gamma=1$). Then, we find the following components for the $Q_{\a\m\n}$ 
%%%%%%%%%%%%%%%%%
\begin{eqnarray}
Q_{000} &=& 0=Q_{0i0}, \qquad Q_{i00}= \frac{2}{c^3}\,\partial_t A_i -
\frac{2}{c^2}\,\partial_i \Phi, \qquad  Q_{00i} =\frac{2}{c^2}\,\aa_i \F, \nonumber \\[2mm]
Q_{ij0} &=& -\frac{1}{c^2}F_{ij} +\frac{1}{c^5}\,\partial_t d_{ij} \qquad
Q_{0ij} = \frac{1}{c^2}\left( \aa_iA_j +\aa_jA_i \right) -\frac{1}{c^5}\,\partial_t d_{ij},
\label{Qamn_alter} \\[2mm]
Q_{i0j} &=& -\frac{F_{ij}}{c^2} -
\frac{2 \aa_t \Phi}{c^3} \h_{ij} +\frac{\partial_t d_{ij}}{c^5},
\,\,\,
Q_{ijk} = -\frac{2 \aa_j \F}{c^2}\h_{ik}
+ \frac{1}{c^4} (\partial_j d_{ik} + \partial_k d_{ij} -\partial_i d_{jk})\,. \nonumber
\end{eqnarray}
%%%%%%%%%%%%%%%%
and $H_{\a\m\n}$ tensors
%%%%%%%%%%%%%%%%%
\begin{eqnarray}
H_{000} &=& 0=H_{00i}, \qquad H_{i00}=\frac{2}{c^2}\,\partial_i \Phi, \qquad  
H_{0i0}  = \frac{2}{c^3}\,\partial_t A_i -\frac{2}{c^2}\,\aa_i \F, \nonumber \\[2mm]
H_{ij0} & =& \frac{1}{c^2}\,F_{ij} -
\frac{2}{c^3}\,\aa_t \Phi\,\h_{ij} +\frac{1}{c^5}\,\partial_t d_{ij}, \qquad
H_{0ij} =-\frac{1}{c^2}\,F_{ij} +\frac{1}{c^5}\,\partial_t d_{ij}, \label{Hamn_alter}   \\[2mm]
H_{i0j} &=&  \frac{1}{c^2} ( \aa_iA_j +\aa_jA_i) 
-\frac{\partial_t d_{ij}}{c^5}\,, \,\,\,
H_{ijk} = -\frac{2 \aa_k \F }{c^2} \h_{ij}
+ \frac{1}{c^4}\,(\partial_k d_{ij} + \partial_i d_{jk} -\partial_j d_{ik})\,. \nonumber 
\end{eqnarray}
%%%%%%%%%%%%%%%%

\end{appendix}



\end{document}